\documentclass[twocolumn,aps,pra,reprint,superscriptaddress,longbibliography]{revtex4-1}

\usepackage{amsmath}
\usepackage{graphicx}
\usepackage[lofdepth,lotdepth, caption=false]{subfig}
\usepackage{verbatim}
\usepackage{color}
\usepackage{dcolumn}
\usepackage{bm}
\usepackage{miller}
\usepackage{color}
\usepackage{xr-hyper}
\usepackage{hyperref}
\usepackage[english]{babel}

\begin{document}

\title{Investigating the magneto-elastic properties in FeSn and Fe$_{3}$Sn$_{2}$ flat band metals} 

\author{Yu Tao}
\affiliation{Department of Physics, University of Virginia, Charlottesville, Virginia 22904, USA}

\author{Luke Daemen}
\affiliation{Neutron Scattering Division, Oak Ridge National Laboratory, Oak Ridge, Tennessee 37831, USA}

\author{Yongqiang Cheng}
\affiliation{Neutron Scattering Division, Oak Ridge National Laboratory, Oak Ridge, Tennessee 37831, USA}

\author{Joerg C.\ Neuefeind}
\affiliation{Neutron Scattering Division, Oak Ridge National Laboratory, Oak Ridge, Tennessee 37831, USA}

\author{Despina Louca}
\thanks{Corresponding author}
\email{louca@virginia.edu}
\affiliation{Department of Physics, University of Virginia, Charlottesville,
Virginia 22904, USA}

\begin{abstract}
Topological quantum magnets FeSn and Fe$_{3}$Sn$_{2}$ were studied using neutron scattering and first-principles calculations. Both materials are metallic but host dispersionless flat bands with Dirac nodes at the $K$ point in reciprocal space. The local structure determined from the pair density function analysis of the neutron diffraction data provided no evidence for electron localization in both compounds, consistent with their metallic nature. At the same time, in FeSn, an anomalous suppression in the $c$-axis lattice constant coupled with changes in the phonon spectra were observed across T$_{N}$ indicating the presence of magneto-elastic coupling and spin-phonon interactions. In addition, it was observed that spin waves persisted well above T$_{N}$, suggesting that the in-plane ferromagnetic spin correlations survive at high temperatures. In contrast, no lattice anomaly was observed in Fe$_{3}$Sn$_{2}$. The inelastic signal could be mostly accounted for by phonons, determined from density functional theory, showing typical softening on warming.
\end{abstract}
\maketitle


\section{Introduction}

Quantum materials are at the forefront of materials research because of their fascinating properties and functionalities that may be harnessed for future applications \cite{tokura_emergent_2017, basov_towards_2017}. The underlying quantum effects, governed by strong interactions and entanglement, give rise to exotic quasiparticles and unconventional states. For example, the competition between the tendency for strong localization via lattice deformations and itinerant behavior manifests itself in textures such as static stripes and polaronic distortions or dynamic charge and spin density wave (CDW and SDW) instabilities, that are prevalent in a wide class of quantum systems \cite{emin_optical_1993, stojchevska_ultrafast_2014, haug_neutron_2010}. In topological materials, the quantum state can be entangled to an extent where its emergent quasiparticles exhibit exotic behaviors that are unique, and cannot be reproduced in conventional solids \cite{yan_topological_2017}. Moreover, these exotic properties are topologically protected as they are robust against symmetry-preserving perturbations \cite{yan_topological_2017, qi_topological_2011, hasan_colloquium_2010}.

One particular feature that has garnered attention recently is the presence of dispersionless flat bands, predicted to play a role in hosting emergent phenomena such as unconventional magnetism and superconductivity \cite{aoki_hofstadter_1996, deng_origin_2003, wu_flat_2007, jacqmin_direct_2014}. A flat band is a macroscopically degenerate manifold of single-particle states with a vanishing electronic bandwidth and quenching of electron kinetic energy. Flat bands have zero curvature resulting to an infinite effective mass and separate electron-like from hole-like bands \cite{mukherjee_observation_2015}. Material candidates with flat bands are twisted graphene, kagome lattices and heavy fermion compounds \cite{tilak_flat_2021, kang_topological_2020, lin_flatbands_2018, ito_high-resolution_1999}.

The kagome lattice consists of a corner-share triangular network of transition metal ions. Due to quantum destructive interference, electrons on kagome nets are localized around the hexagons \cite{kang_dirac_2020, li_realization_2018}. Dispersionless electronic flat bands have been observed in the metallic kagome lattice materials of FeSn, Fe$_{3}$Sn$_{2}$ and Co$_{3}$Sn$_{2}$S$_{2}$, determined from angle-resolved photoemission spectroscopy (ARPES) measurements \cite{kang_dirac_2020, lin_flatbands_2018, xu_electronic_2020}. Some kagome lattice materials also have nontrivial topologically protected bands. Recent examples include FeSn, Fe$_{3}$Sn$_{2}$ and YMn$_{6}$Sn$_{6}$, where Dirac points have been observed \cite{kang_dirac_2020, ye_massive_2018, li_dirac_2021}, and Co$_{3}$Sn$_{2}$S$_{2}$, a ferromagnetic (FM) Weyl semimetal with a giant anomalous Hall effect \cite{xu_electronic_2020}. Kagome lattice materials are also an ideal platform to explore geometrically frustrated magnetism. In herbertsmithite, the kagome nets lead to frustration of the long-range antiferromagnetic (AFM) order and result in a spin-liquid state \cite{han_fractionalized_2012}. Other properties observed in kagome lattice materials include CDW and superconductivity reported in AV$_{3}$Sb$_{5}$ (A = Rb, Cs) \cite{li_observation_2021}.

\begin{figure}[h]
\begin{center}
\includegraphics[width=8.6cm]
{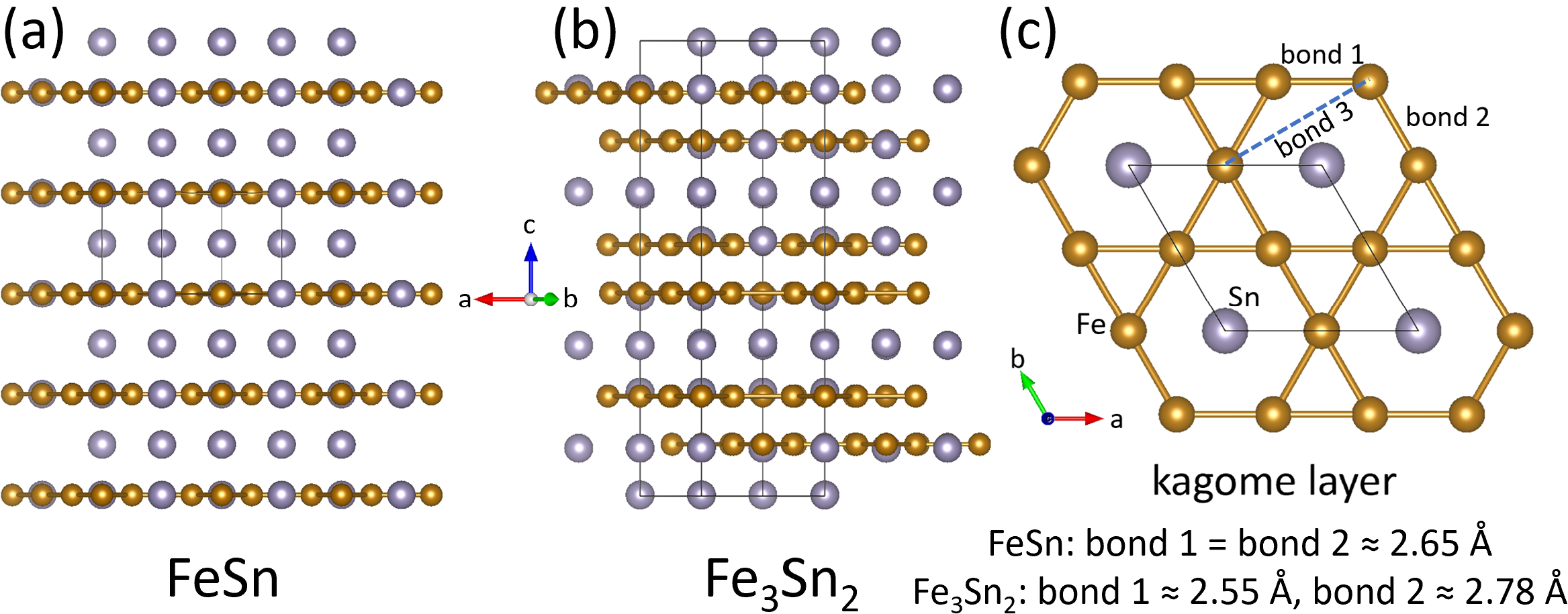}
\end{center}
\caption{Crystal structures of kagome metals FeSn and Fe$_{3}$Sn$_{2}$. The layer stacking orders along the $c$-axis for FeSn and Fe$_{3}$Sn$_{2}$ are shown in panel (a) and (b), respectively. The in-plane geometry of a single kagome layer of Fe$_{3}$Sn is shown in panel (c). Solid black lines indicate the size of the unit cell. For FeSn, bond 1 length = bond 2 length $\approx$ 2.65 $\AA$. For Fe$_{3}$Sn$_{2}$, bond 1 length $\approx$ 2.55 $\AA$, bond 2 length $\approx$ 2.78 $\AA$.}
\label{fig:fig1}
\end{figure}

In this work, the kagome lattice materials FeSn (space group $P6/mmm$) and Fe$_{3}$Sn$_{2}$ (space group $R\overline{3}m$) were studied. Both systems were reported to exhibit flat bands and Dirac points in their electronic band structures \cite{kang_dirac_2020, lin_flatbands_2018, ye_massive_2018}. Their crystal structure consists of kagome layers of Fe$_{3}$Sn stacked along the $c$-axis with layers of pure Sn in between, as shown in Fig.\ \ref{fig:fig1}(a,b). The difference between the two compounds is that in FeSn, layers of Fe$_{3}$Sn alternate with Sn layers, and all the Fe-Fe bonds in the kagome nets have the same bond length, but in Fe$_{3}$Sn$_{2}$, bilayers of Fe$_{3}$Sn alternate with Sn layers, and two types of Fe-Fe bonds are found in a single kagome layer (bond 1 and bond 2 as shown in Fig.\ \ref{fig:fig1}(c)) \cite{tanaka_three-dimensional_2020}.

The stacking order is responsible for the complex magnetic behavior arising from competing magnetic interactions and geometrical frustration. FeSn exhibits long-range AFM order below T$_{N}$ = 365 K, where spins align ferromagnetically in-plane but antiferromagnetically out-of-plane \cite{yamaguchi_neutron_1967, haggstrom_studies_1975, sales_electronic_2019}. Fe$_{3}$Sn$_{2}$ has a FM structure with a T$_{C}$ $\approx$ 660 K and the spins aligned in the $c$-direction. It has been suggested that a spin reorientation transition (SRT) occurs at around 100 K, where the spins flip to in-plane \cite{malaman_magnetic_1978, caer_magnetic_1979, fenner_non-collinearity_2009}. In addition, studies on low-energy spin waves in Fe$_{3}$Sn$_{2}$ using small-angle inelastic neutron scattering indicate that Fe$_{3}$Sn$_{2}$ is an isotropic ferromagnet to at least 480 K \cite{dally_isotropic_2021}.

Flat bands can enable a transition to a topological phase that contains two Weyl nodes in 3-dimensional (3D) systems \cite{lu_weyl_2013}. This occurs by applying spin-orbit coupling (SOC) and time-reversal symmetry (TRS) breaking. Traditionally, the mechanism by which Weyl nodes are formed requires either breaking TRS or inversion symmetry of a Dirac point. However, in 3D systems with flat bands, under strong SOC, the dispersionless bands separate from the dispersive bands. When further band splitting is induced by magnetization, a much larger gap than the SOC gap of the intermediate phase occurs \cite{yin_negative_2019}. Thus in the presence of both SOC and Zeeman splitting, a 3D flat band system becomes a Weyl semimetal while breaking the macroscopic degeneracy. If magnetization is stronger than SOC, at a minimum, one would expect a single pair of Weyl nodes.

Though the electronic band structures of FeSn and Fe$_{3}$Sn$_{2}$ have been extensively investigated, little is known of the phonon and magnetic dynamics. Structural distortions are often seen in kagome lattice materials. For example, the CDW state of CsV$_{3}$Sb$_{5}$ is associated with a static lattice distortion that results in a 2×2×2 superlattice structure below the transition temperature of 94 K, which is also accompanied by a sudden hardening of a longitudinal optical phonon mode \cite{xie_electron-phonon_2022}. Since a flat band dispersion in reciprocal space is due to strong electron localization in real space, it is expected to introduce distortions in the lattice due to localization.

Moreover, it was recently shown in FeSn that a large electron-magnon interaction is present at 5 K, which results in strong damping of the high energy magnon spectra, due to interactions with the Stoner continuum \cite{do_damped_2022, xie_spin_2021}. The reported magnon results emphasize the role of itinerant carriers on the topological spin excitations of metallic kagome magnets. Furthermore, a change in the magnetism might have direct consequences on its phonon dynamics. In Fe$_{3}$Sn$_{2}$, Raman spectroscopy showed an anomaly in the linewidth for one of the A$_{g}$ modes at 100 K, which might be due to the SRT \cite{wu_evidence_2021}. On the other hand, no phonon measurements have been carried out on FeSn. Examples of other kagome lattice materials that have changes in lattice constants and phonons which accompany changes in spin correlations include Na$_{2}$Ti$_{3}$Cl$_{8}$ and Dy$_{3}$Ru$_{4}$Al$_{12}$ \cite{paul_spin-lattice_2020, henriques_magneto-elastic_2016}.

To explore the interplay between spin and lattice degrees of freedom and the associated phonon behaviors in FeSn and Fe$_{3}$Sn$_{2}$, neutron scattering measurements were performed. In FeSn, an unusual $c$-axis lattice constant anomaly was observed between 300 and 500 K, across its Néel temperature, indicating magneto-elastic coupling is likely to present. No evidence of local structure distortions was observed in FeSn and Fe$_{3}$Sn$_{2}$ due to the strong electron localization that results in the flat bands. Despite the AFM transition, spin waves in FeSn persist on warming to at least 450 K, suggesting in-plane FM ordering that forms well above T$_{N}$. In addition, changes in the phonon softening behavior were observed in FeSn across T$_{N}$ but not in Fe$_{3}$Sn$_{2}$, indicating spin-phonon coupling in FeSn.


\section{Experimental and Calculation Details}

FeSn and Fe$_{3}$Sn$_{2}$ powders were prepared using solid state reaction. For FeSn, elemental powders of Fe and Sn, in a mole ratio of 1:1, were thoroughly mixed and pressed into a pellet. The pellet was then sealed into an evacuated silica ampoule and heated at 670 $^{\circ}$C for one to two days. For Fe$_{3}$Sn$_{2}$, a stoichiometric ratio of Fe and Sn powders was used, and the sintering was done in an evacuated quartz silica ampoule at 770 $^{\circ}$C for approximately a week, followed by quenching in water.

The neutron scattering experiments were carried out on FeSn and Fe$_{3}$Sn$_{2}$ powders on the time-of-flight spectrometer VISION and diffractometer NOMAD at the Spallation Neutron Source (SNS) of Oak Ridge National Laboratory (ORNL). For the measurements on VISION, 10 g each of FeSn and Fe$_{3}$Sn$_{2}$ powders were used and elastic and inelastic data were collected simultaneously. The inelastic neutron scattering intensity was measured along two narrow paths in ($Q$, $E$) space, which are labeled as the low-$Q$ and high-$Q$ trajectories. These trajectories are determined by the fixed final neutron energy of 3.5 meV and the scattering angles of 45$^{\circ}$ and 135$^{\circ}$. Background from an empty vanadium can scan was subtracted from all data. For the measurements on NOMAD, 2 g each of FeSn and Fe$_{3}$Sn$_{2}$ were sealed into vanadium cans and measured at 92 and 290 K for 3 hours per scan. The pair distribution function (PDF) analysis was performed using the NOMAD neutron diffraction data. Neutron diffraction measurements on NOMAD allow acquisition of the total scattering function $S(Q)$ at very high momentum transfers. The real-space atomic pair distribution function $G(r)$ can then be obtained by Fourier transforming the $S(Q)$ data. PDF analysis provides quantitative insight into the local structure of materials where the structural correlations extend only over a few angstroms.

Spin-polarized Density Functional Theory (DFT) calculations of FeSn and Fe$_{3}$Sn$_{2}$ were performed using the Vienna Ab initio Simulation Package (VASP) \cite{kresse_efficient_1996}. The calculation used Projector Augmented Wave (PAW) method \cite{blochl_projector_1994, kresse_ultrasoft_1999} to describe the effects of core electrons, and Perdew-Burke-Ernzerhof (PBE) \cite{perdew_generalized_1996} implementation of the Generalized Gradient Approximation (GGA) for the exchange-correlation functional. The energy cutoff was 600 eV for the plane-wave basis of the valence electrons. The reported lattice parameters and atomic coordinates were used as the initial input for the structure \cite{van_der_kraan_57fe_1986, malaman_structure_1976}, and were fully relaxed to minimize the potential energy. The electronic structure of the magnetic primitive cell was calculated on a $\Gamma$-centered mesh (9×9×5 for FeSn and 15×15×15 for Fe$_{3}$Sn$_{2}$). The total energy tolerance for electronic energy minimization was 10$^{-8}$ eV, and for structure optimization was 10$^{-7}$ eV. The maximum interatomic force after relaxation was below 0.001 eV/$\AA$. A supercell (3×3×1 for FeSn and 2×2×2 for Fe$_{3}$Sn$_{2}$) was created for phonon calculations. The interatomic force constants were calculated by Density Functional Perturbation Theory (DFPT), and the vibrational eigen-frequencies and modes as well as the (partial) phonon density of states (DOS) were then calculated using Phonopy \cite{togo_first_2015}. The OCLIMAX software \cite{cheng_simulation_2019} was used to convert the DFT-calculated phonon results to the simulated VISION spectra (phonon contribution).


\section{Results and Discussion}
\subsection{Static structures}

\begin{figure}[h]
\begin{center}
\includegraphics[width=8.6cm]
{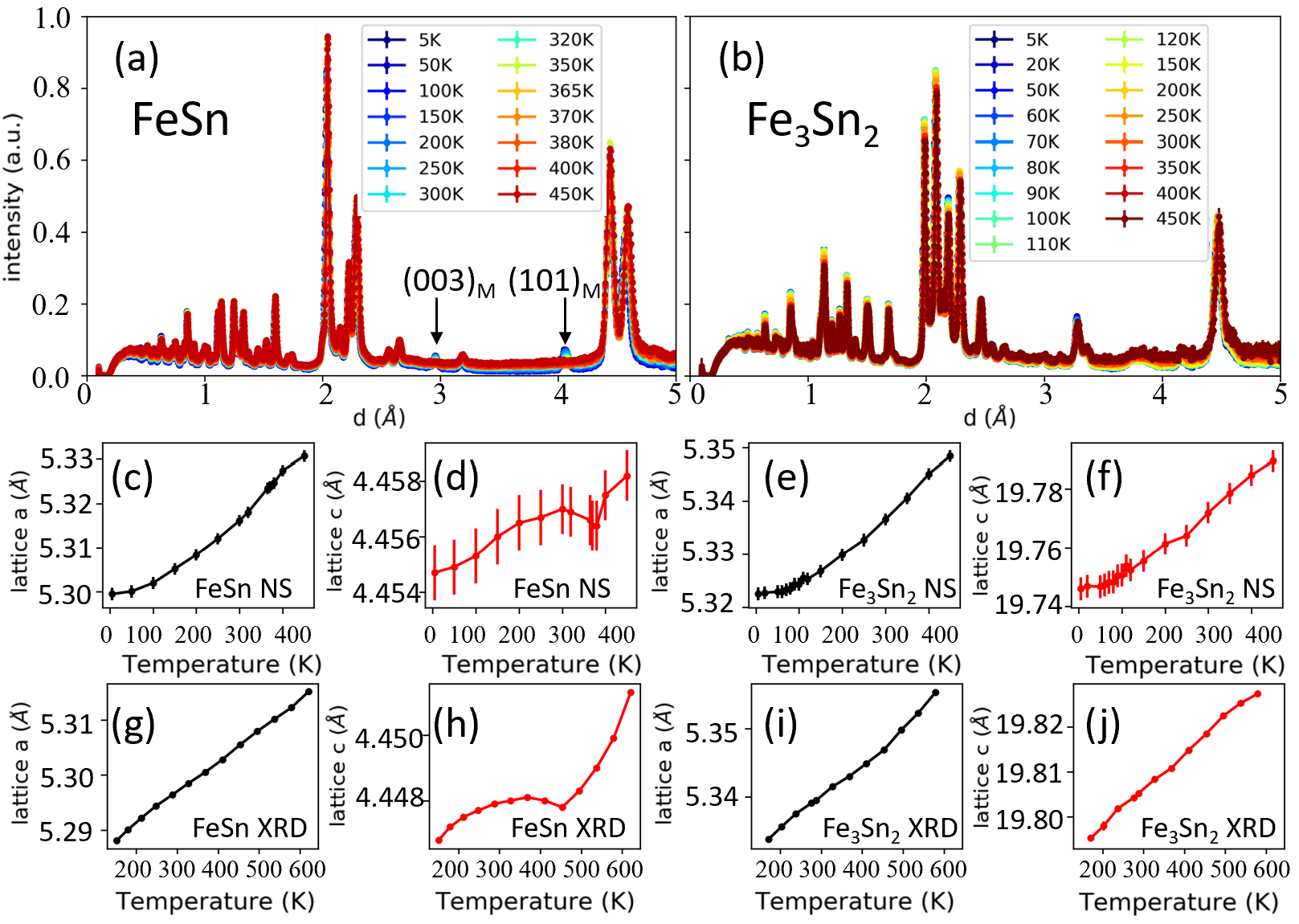}
\end{center}
\caption{(a,b) Elastic neutron scattering intensity measured on (a) powder FeSn and (b) powder Fe$_{3}$Sn$_{2}$ as a function of temperature on warming from 5 to 450 K. (c-f) $a$ and $c$ lattice constants for (c,d) FeSn and (e,f) Fe$_{3}$Sn$_{2}$ as a function of temperature from VISION neutron scattering (NS) data. (g-j) $a$ and $c$ lattice constants for (g,h) FeSn and (i,j) Fe$_{3}$Sn$_{2}$ as a function of temperature from XRD data.}
\label{fig:fig2}
\end{figure}

Shown in Fig.\ \ref{fig:fig2}(a,b) are the elastic neutron scattering intensity measured on FeSn and Fe$_{3}$Sn$_{2}$ as a function of temperature between 5 and 450 K. The data were plotted in $d$-spacing. No structural transition is observed and the diffraction patterns at all temperatures can be fit using the crystal symmetry. The $a$ and $c$ lattice constants of FeSn and Fe$_{3}$Sn$_{2}$, obtained from the Rietveld refinement on the neutron scattering data shown in Fig.\ \ref{fig:fig2}(a,b) are plotted in Fig.\ \ref{fig:fig2}(c-f) as a function of temperature. In FeSn, an anomalous reduction of the $c$ lattice constant is observed on warming between 300 and 400 K (Fig.\ \ref{fig:fig2}(d)), whereas the $a$ lattice constant shows a usual thermal expansion (Fig.\ \ref{fig:fig2}(c)). In contrast, Fe$_{3}$Sn$_{2}$ shows a uniform expansion in both the $a$ and $c$ lattice constants on warming from 5 to 450 K, which is below its Curie temperature.

The $c$-axis anomaly observed in FeSn was also observed in the X-ray diffraction (XRD) measurements performed between 150 and 650 K. The lattice parameters determined from the Rietveld refinement are plotted in Fig.\ \ref{fig:fig2}(g-j). From Fig.\ \ref{fig:fig2}(h), it is observed that the $c$ lattice constant of FeSn first increases upon warming at low temperatures, and it plateaus between 300 and 450 K. Upon further warming, the lattice constant increases roughly linearly with temperature again. The anomaly in FeSn occurs across the Néel temperature of 365 K, suggesting that the interplane atomic correlations might be coupled with the magnetism, due to magneto-elastic coupling.

\begin{figure}[h]
\begin{center}
\includegraphics[width=8.6cm]
{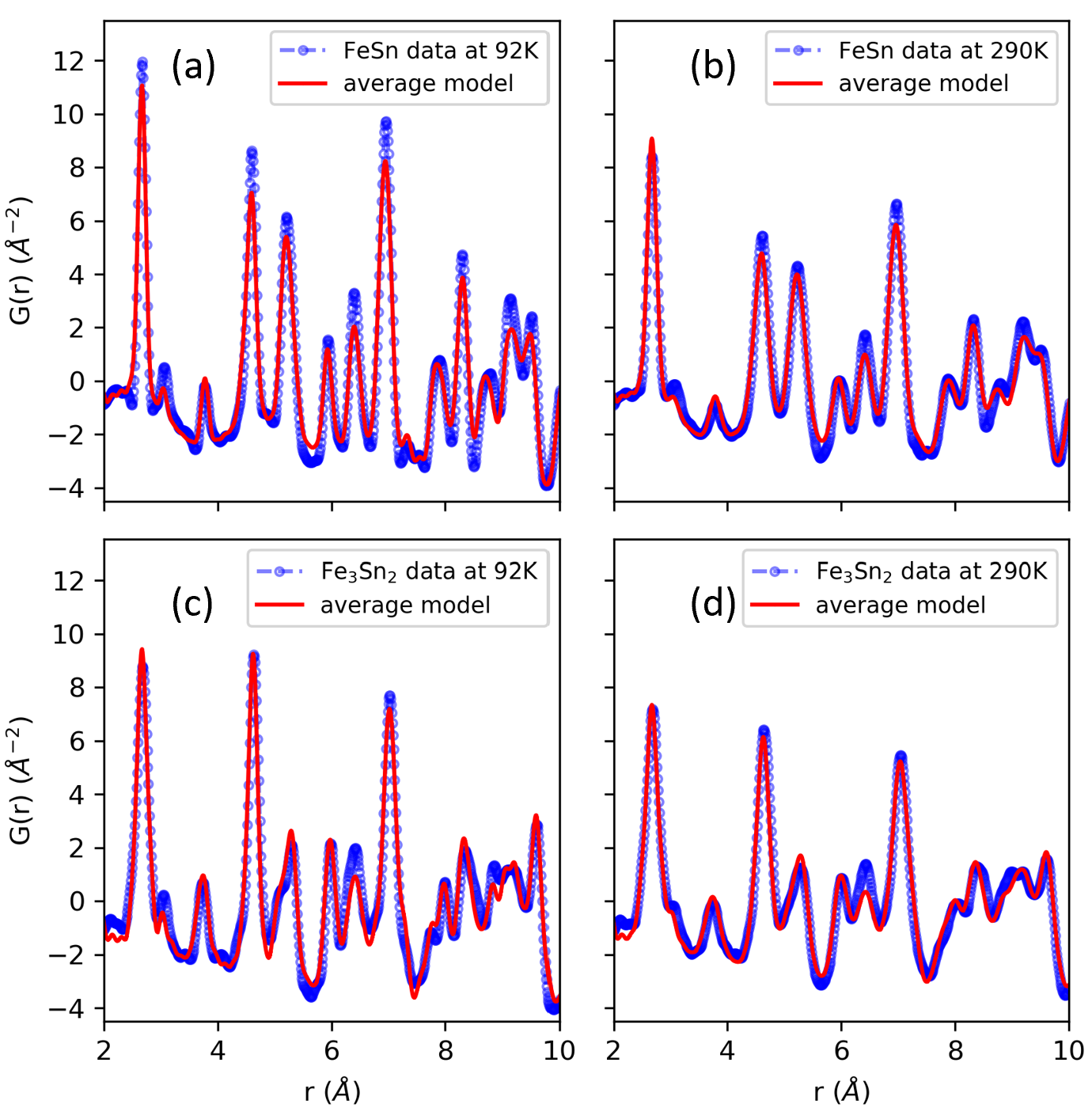}
\end{center}
\caption{(a,b) The PDF of the local structure of FeSn at 92 and 290 K (dashed blue lines), compared with the average model (solid red lines). (c,d) The PDF of the local structure of Fe$_{3}$Sn$_{2}$ at 92 and 290 K (dashed blue lines), compared with the average model (solid red lines).}
\label{fig:fig3}
\end{figure}

Shown in Fig.\ \ref{fig:fig3} are the $G(r)$ up to 10 $\AA$ for FeSn and Fe$_{3}$Sn$_{2}$ at 92 and 290 K. The PDF data (blue symbols) agrees well with a model $G(r)$ (red curves) calculated using the atomic coordinates and unit cell dimensions from Ref.\cite{van_der_kraan_57fe_1986, malaman_structure_1976}, with the proposed $P6/mmm$ symmetry for FeSn, and $R\overline{3}m$ symmetry for Fe$_{3}$Sn$_{2}$. No deviations of the local structure is observed from the average symmetry that would indicate electron lattice coupling. Given the metallic nature of both systems, electron-lattice interactions in FeSn and Fe$_{3}$Sn$_{2}$ are most likely weak, which can also be understood from the fact that the electronic bands are flat in only a limited region in reciprocal space (along $\Gamma$-K-M below the Fermi level at about -0.2 eV for both materials \cite{kang_dirac_2020, lin_flatbands_2018} and the energy of the flat bands is too high and does not couple with the lattice.

\begin{figure}[h]
\begin{center}
\includegraphics[width=8.6cm]
{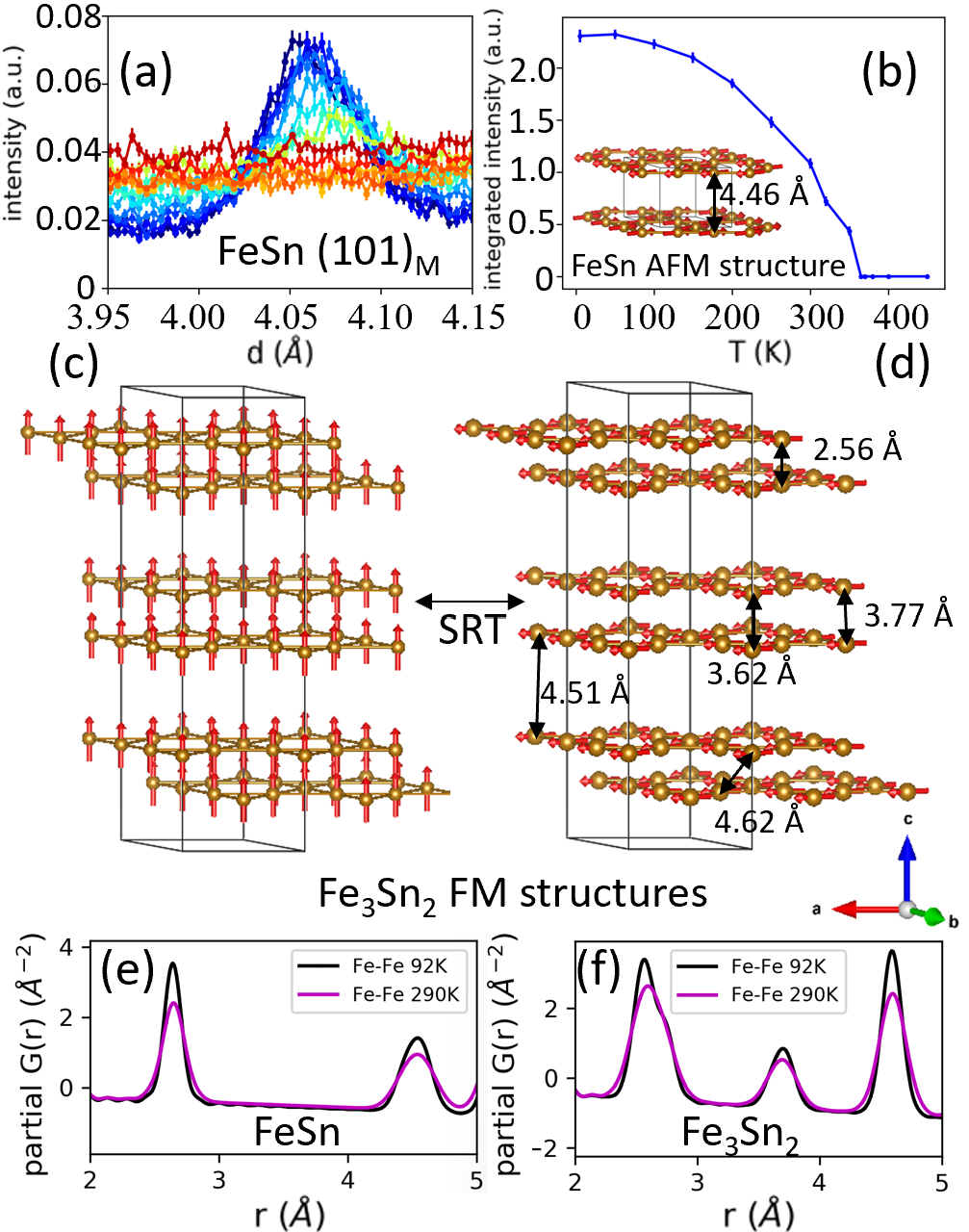}
\end{center}
\caption{(a) Elastic neutron scattering intensity tracking the (101)$_{M}$ AFM Bragg peak of FeSn as a fuction of temperature. (b) Integrated intensity of the (101)$_{M}$ AFM Bragg peak as a function of temperature, showing the Néel transition at around 365 K. Inset: AFM structure of FeSn, with spins ferromagnetically aligned in-plane and antiferromagnetically coupled along the $c$-axis. (c,d) FM structures of Fe$_{3}$Sn$_{2}$. A SRT occurs at $\sim$ 100 K on cooling where the spins reorientate from (c) to (d). (e,f) Partial PDFs that show only the Fe-Fe correlations in FeSn and Fe$_{3}$Sn$_{2}$ up to 5 $\AA$.}
\label{fig:fig4}
\end{figure}

Two AFM Bragg peaks, (003)$_{M}$ and (101)$_{M}$, were observed in FeSn from Fig.\ \ref{fig:fig2}(a). Shown in Fig.\ \ref{fig:fig4}(a) is the (101)$_{M}$ AFM Bragg peak in FeSn zoomed in from Fig.\ \ref{fig:fig2}(a). It is plotted as a function of temperature, and its integrated intensity is shown in Fig.\ \ref{fig:fig4}(b). The decrease of the magnetic peak intensity on warming leads to a continuous increase in the background as seen in the figure. This might be due to paramagnetic scattering arising from fluctuation of disordered spins. The (101)$_{M}$ peak disappears at T$_{N}$. For the FM Fe$_{3}$Sn$_{2}$, though the positions of the magnetic Bragg peaks overlap with those of the nuclear Bragg peaks, a decrease of the Bragg peak intensity upon warming, due to an increase in spin fluctuations, can still be seen in Fig.\ \ref{fig:fig2}(b).

The AFM structure of FeSn is shown in the inset of Fig.\ \ref{fig:fig4}(b). The high- and low-temperature FM structures of Fe$_{3}$Sn$_{2}$ are shown in Fig.\ \ref{fig:fig4}(c,d), where a SRT occurs at $\sim$100 K. The difference in magnetism between the two compounds is related to the different magnetic exchange interactions between the Fe inter-planes. Of the five intermetallic Fe-Sn materials that have been studied so far (FeSn$_{2}$, FeSn, Fe$_{3}$S$_{2}$, Fe$_{5}$Sn$_{3}$ and Fe$_{3}$Sn), when the mole ratio of Fe to Sn is smaller or equal to 1, such as in FeSn and FeSn$_{2}$ \cite{kanematsu_antiferromagnetism_1960}, the spin ordering is AFM. In Fe$_{3}$S$_{2}$, Fe$_{5}$Sn$_{3}$ \cite{li_large_2020} and Fe$_{3}$Sn \cite{sales_ferromagnetism_2014}, the magnetic coupling is FM. In the AFM compounds, an Sn layer separates neighboring Fe layers. In the FM compounds, the Fe layers are much closer; Fe$_{3}$Sn consists only of Fe$_{3}$Sn kagome layers stacked along $c$, and Fe$_{5}$Sn$_{3}$ has pure Fe layers and Fe-Sn layers (not kagome) alternate along the $c$-axis \cite{giefers_high_2006}.

The Sn layers that modulate the magnetic structures of FeSn and Fe$_{3}$Sn$_{2}$ changes the bond distances between the inter-plane Fe atoms and it can be understood from the PDF data. Shown in Fig.\ \ref{fig:fig4}(e,f) are the partial PDFs corresponding to Fe-Fe pair correlations in FeSn and Fe$_{3}$Sn$_{2}$ up to 5 $\AA$, which were obtained from the total PDF shown in Fig.\ \ref{fig:fig3}, since the total PDF is a superposition of partial PDFs corresponding to correlations between all the atoms in the sample. For FeSn, the first peak is due to the nearest Fe-Fe neighbor in-plane (bond 1 and bond 2 at $\sim$2.6 $\AA$ as shown in Fig.\ \ref{fig:fig1}(c)) and the second peak comes from the second nearest Fe-Fe pairs in-plane (bond 3 at $\sim$4.6 $\AA$ as shown in Fig.\ \ref{fig:fig1}(c)) and the nearest Fe-Fe pairs inter-plane ($\sim$4.5 $\AA$ as shown in Fig.\ \ref{fig:fig4}(b)). In contrast, three partial PDF peaks are present in Fe$_{3}$Sn$_{2}$ up to 5 $\AA$. The first peak is from the nearest Fe-Fe neighbor in-plane (bond 1 and bond 2 as shown in Fig.\ \ref{fig:fig1}(c), at $\sim$2.6 $\AA$ and $\sim$2.8 $\AA$) and the nearest Fe-Fe neighbor inter-plane ($\sim$2.6 $\AA$ as shown in Fig.\ \ref{fig:fig4}(d)). The second peak comes from the second nearest Fe-Fe pairs inter-plane ($\sim$3.6 $\AA$ and $\sim$3.8 $\AA$ as shown in Fig.\ \ref{fig:fig4}(d)), and the third peak is from the second nearest Fe-Fe pairs in-plane (bond 3 at $\sim$4.6 $\AA$) and additional inter-plane Fe-Fe bonds ($\sim$4.5 $\AA$ and $\sim$4.6 $\AA$ as shown in Fig.\ \ref{fig:fig4}(c,d)). Therefore, more short-range inter-plane magnetic exchange interactions are expected for Fe$_{3}$Sn$_{2}$ than FeSn. Given Fe atoms tend to align ferromagnetically (e.g. $\alpha$-Fe is FM below 770 $^{\circ}$C), it would cost the Fe$_{3}$Sn$_{2}$ system too much energy to align antiferromagnetically given that the Fe pairs are too close together.

\subsection{Phonon and magnon dynamics}

\begin{figure}[h!]
\begin{center}
\includegraphics[width=8.6cm]
{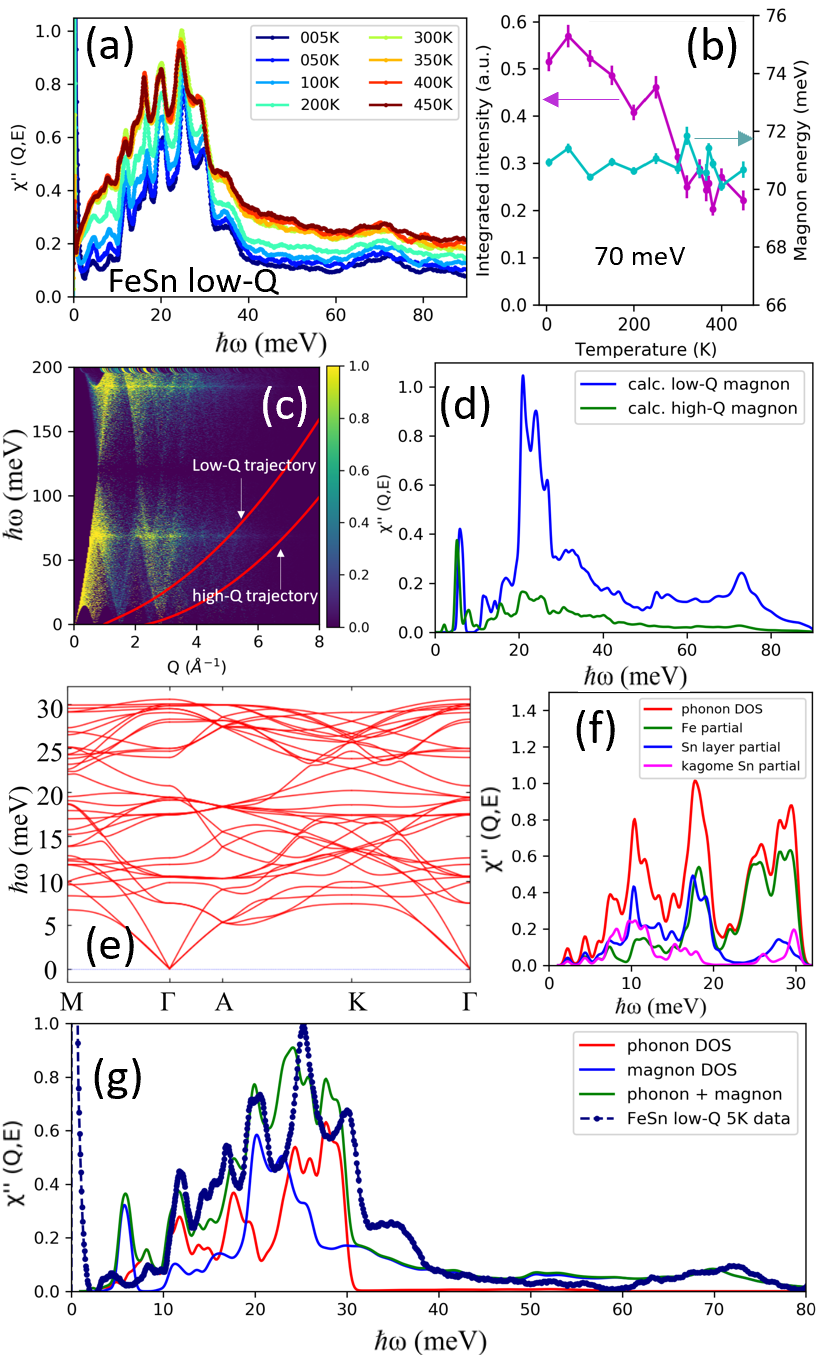}
\end{center}
\caption{(a) Dynamic susceptibility $\chi^{\prime \prime}(Q, E)$ for FeSn collected along the low-$Q$ trajectory as a function of temperature. (b) Position and integrated intensity of the magnon peak at $\sim$70 meV. (c) Simulated powder averaged spin waves for FeSn. Red lines indicate the low-$Q$ and high-$Q$ trajectories. The Dirac point is at $\sim$120 meV. (d) Simulated magnon DOS for FeSn along the low-$Q$ and high-$Q$ trajectories. (e) DFT calculation of the phonon dispersions for FeSn. (f) Calculated phonon DOS for FeSn and partial phonon DOS from Fe, Sn in the kagome layer and Sn in the Sn layer. (g) Calculated magnon DOS and phonon DOS along the low-$Q$ trajectory, compared with the FeSn data at 5 K along the low-$Q$ trajectory.}
\label{fig:fig5}
\end{figure}

The inelastic neutron scattering data collected at VISION provide both phonon and magnon DOS information. Shown in Fig.\ \ref{fig:fig5}(a) is the dynamic susceptibility (Bose-factor-corrected inelastic neutron scattering intensity), $\chi^{\prime \prime}(Q, E)$, for FeSn along the low-$Q$ trajectory. It is plotted as a function of temperature, with the data shifted along the $y$-axis for clarity. Several changes are observed in $\chi^{\prime \prime}(Q, E)$ that may indicate changes in phonon and magnon modes. Fig.\ \ref{fig:fig5}(c) is a plot of the spin waves at 5 K calculated using spinW \cite{toth_linear_2015}. Linear spin-wave theory (LSWT) with the Hamiltonian, $H$ = $J_{n} \sum_{<i,j>} S_{i} \cdot S_{j} - D_{z} \sum_{i} (S_{i}^{z})^{2}$, was used. The reported Heisenberg magnetic coupling exchange constants were obtained from Ref.\ \cite{do_damped_2022} that consider up to the fourth nearest neighbor in-plane coupling and the second nearest neighbor inter-plane coupling. Previous single crystal neutron scattering measurements showed that spin waves in FeSn extend to well above 200 meV and the Dirac magnons are reported to appear at $\sim$120 meV \cite{do_damped_2022, xie_spin_2021}. Both features were reproduced in the simulation. Also shown in Fig.\ \ref{fig:fig5}(c) are the two trajectories across which data are collected on VISION. Only low energy magnons are sampled along the low- and high-$Q$ trajectories. The calculated powder averaged spin wave intensity was integrated along the two trajectories and convoluted with the instrument resolution function. The calculated magnon DOS along the two trajectories are shown in Fig.\ \ref{fig:fig5}(d). The magnon DOS along the high-$Q$ trajectory is much weaker than the one along the low-$Q$ trajectory due to the magnetic form factor of Fe.

From the calculated magnon DOS, it is observed that a $\sim$70 meV peak observed in the data is magnetic in nature. In Fig.\ \ref{fig:fig5}(b), the integrated intensity of this peak is plotted as a function of temperature. Even though the intensity decreases on warming, it does not go to zero. Instead it persists well above T$_{N}$. At the same time, the position of the peak remains essentially unchanged with increasing temperature through T$_{N}$. One possible explanation why the magnons persist above T$_{N}$ might be due to residual short-range magnetic ordering related to geometrical frustration in the kagome antiferromagnet. Similarly, in the frustrated triangular antiferromagnet YMnO$_{3}$, magnetic excitations have also been observed above the Néel temperature from inelastic neutron scattering measurements \cite{demmel_persistent_2007}. 

Based on the spin wave simulations, there should be a magnon peak at $\sim$5 meV, however, no such peak is seen in the data. This discrepancy could be due to peak broadening from the $Q$-dependent resolution of the instrument that was not taken into account in the model. During the calculation for magnon DOS, only an energy-dependent resolution function for the VISION instrument was implemented, which is approximately described by a quadratic equation reported in Ref. \cite{cheng_simulation_2019}. The resolution along Q would further smooth out the $\sim$5 meV peak, thus better reproducing the data.

DFT calculations were also carried out to obtain the phonon dispersions for FeSn as seen in Fig.\ \ref{fig:fig5}(e). Fig.\ \ref{fig:fig5}(f) is a plot of the calculated phonon DOS and partial phonon DOS for Fe, Sn in the kagome layer, and Sn in the Sn layer. The calculated phonon DOS is consistent with that reported in Ref.\ \cite{ptok_chiral_2021}. The DFT calculations suggest that phonon modes in FeSn are present between 0 and $\sim$30 meV \cite{ptok_chiral_2021}. Thus, the inelastic peak at $\sim$70 meV in Fig.\ \ref{fig:fig5}(a) is clearly from spin waves.

Fig.\ \ref{fig:fig5}(g) is a comparison of the total magnon + phonon DOS with $\chi^{\prime \prime}(Q, E)$. The calculated magnon DOS was first scaled using the 70 meV magnon peak, and the calculated phonon DOS was then adjusted accordingly to match the overall intensity. Note that the intensity of the calculated phonon DOS in Fig.\ \ref{fig:fig5}(g) is different from that in Fig.\ \ref{fig:fig5}(f), due to an intensity correction in order to simulate the VISION spectra. The calculated total inelastic neutron scattering intensity overall agrees with the data at 5 K along the low-$Q$ trajectory. In addition to the 70 meV magnon peak, the inelastic peak at $\sim$35 meV is also likely from spin waves. The inelastic peak at $\sim$20 meV mainly comes from spin waves as well, with a small contribution from phonons. The magnetic origin of this peak is also evident from the intensity along the low-$Q$ trajectory, which is larger than that along the high-$Q$ trajectory, which is what one would expect from the $Q$-dependence of the magnetic form factor. Meanwhile, phonons contribute to inelastic peaks at around 8, 12, 17, 25 and 30 meV.

\begin{figure}[h]
\begin{center}
\includegraphics[width=8.6cm]
{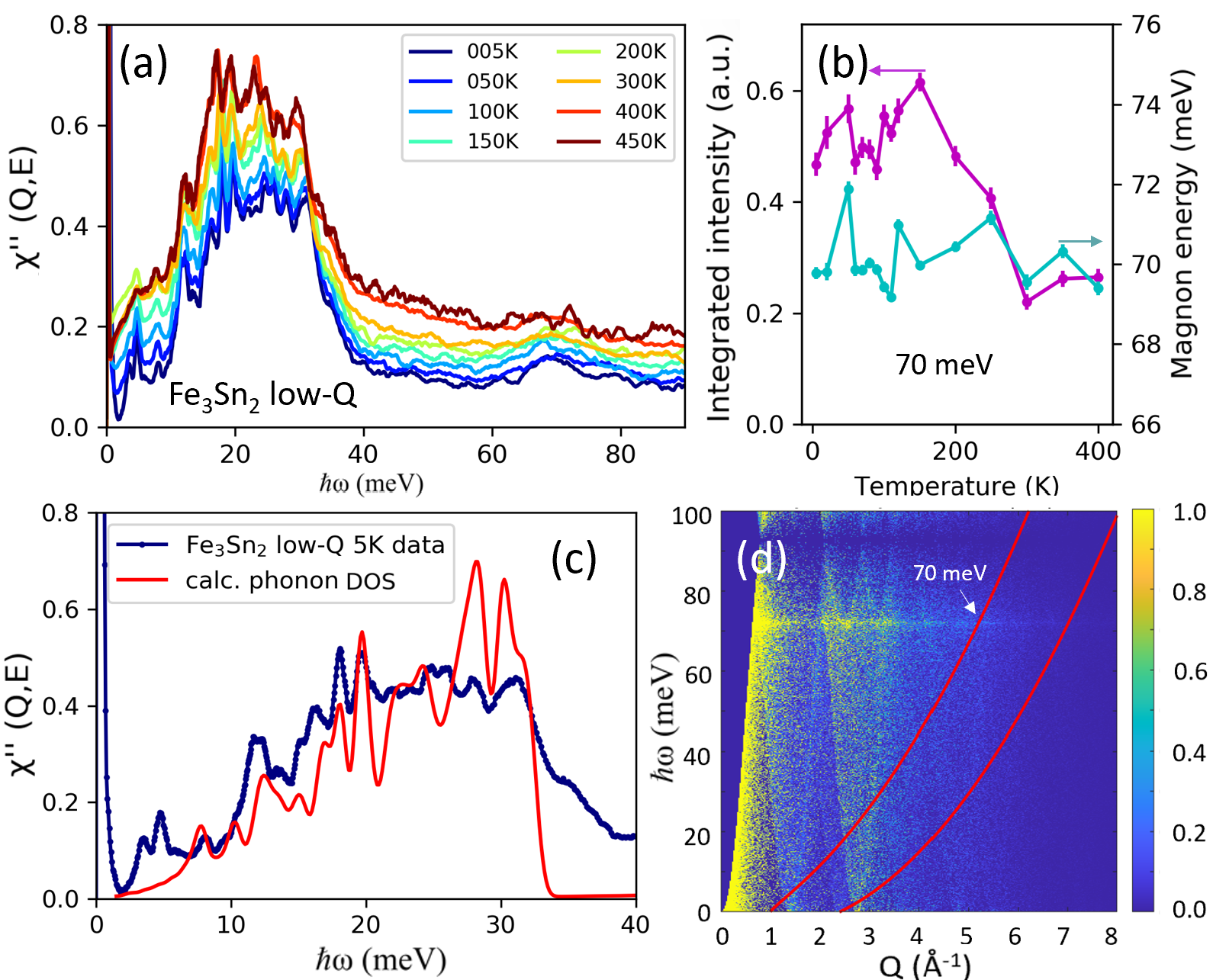}
\end{center}
\caption{(a) Dynamic susceptibility $\chi^{\prime \prime}(Q, E)$ for Fe$_{3}$Sn$_{2}$ collected along the low-$Q$ trajectory as a function of temperature. (b) Energy and integrated intensity of the magnon peak at $\sim$ 70 meV. (c) DFT calculation of the phonon DOS for Fe$_{3}$Sn$_{2}$ compared with the 5 K data along the low-$Q$ trajectory. (d) Simulated powder averaged spin waves for Fe$_{3}$Sn$_{2}$ using a simple planer FM model. Red lines indicate the low-$Q$ and high-$Q$ trajectories.}
\label{fig:fig6}
\end{figure}

Similar analysis was performed on the low-$Q$ trajectory inelastic neutron scattering data collected on Fe$_{3}$Sn$_{2}$ to study its spin wave and phonon behaviors. Shown in Fig.\ \ref{fig:fig6}(a) is the dynamic susceptibility $\chi^{\prime \prime}(Q, E)$ along the low-$Q$ trajectory as a function of temperature, plotted with the data shifted along the $y$-axis for clarity. Just like in FeSn, an inelastic peak is observed at $\sim$70 meV that persists through all the temperatures up to 450 K. The temperature dependence of the integrated intensity and position of this peak are shown in Fig.\ \ref{fig:fig6}(b). The 70 meV peak is due to spin waves and all measurements were carried out below the Curie temperature 660 K. Moreover, DFT calculations were performed on the FM phase to obtain the phonon DOS, and the result is shown in Fig.\ \ref{fig:fig6}(c). Phonons in Fe$_{3}$Sn$_{2}$ extend to $\sim$33 meV, which suggests that inelastic neutron scattering intensity above 33 meV is likely from the magnetic spectra. Unlike in FeSn, little is known of the magnetic interactions in Fe$_{3}$Sn$_{2}$, and due to the difficulty in finding enough magnon peaks that are isolated from the phonon peaks from the powder data, the magnon DOS could not be accurately simulated. Nevertheless, by comparing only the calculated phonon DOS with the 5 K data in Fig.\ \ref{fig:fig6}(c), a qualitative agreement between the two can be seen in certain regions, such as at 8 meV and between 20 and 25 meV. Thus the first two inelastic peaks at $\sim$3 and 5 meV are probably from the magnon, and that the spin waves in Fe$_{3}$Sn$_{2}$ might be relatively weak where it overlaps with the phonons.

In-plane FM interactions might be responsible for the 70 meV magnon peak in both FeSn and Fe$_{3}$Sn$_{2}$. For FeSn, from Fig.\ \ref{fig:fig4}(a), the intensity of the magnetic Bragg peaks decreases on warming and disappears by T$_{N}$, indicating the disappearance of the AFM spin order. On the other hand, the 70 meV magnon peak persists well above T$_{N}$ up to at least 450 K, suggesting the existence of non-paramagnetic spin correlations. One possible explanation for this seemingly contradictory magnetic behavior in FeSn is that the 70 meV magnon peak is due to remaining in-plane FM interactions, since ferromagnetism in Fe$_{3}$Sn$_{2}$ also produces a magnon peak at a similar position.

To support this assumption, additional spin wave simulations were performed for Fe$_{3}$Sn$_{2}$. Shown in Fig.\ \ref{fig:fig6}(d) is the calculated powder averaged spin waves using the FM structure of Fe$_{3}$Sn$_{2}$ at low temperature (Fig.\ \ref{fig:fig4}(e)), but with the reported in-plane magnetic exchange coupling constants for FeSn. This is because both compounds share a very similar in-plane kagome lattice geometry. No inter-plane magnetic couplings was added. Though the simulation is not an accurate representation of the actual magnetic interactions in Fe$_{3}$Sn$_{2}$, this simple planar FM structure does produce a spin wave intensity at $\sim$70 meV where the low-$Q$ trajectory cuts through (indicated by the white arrow in Fig.\ \ref{fig:fig6}(d)). Hence, the 70 meV magnon peak is likely to be FM in origin.

Given the reported magnetic exchange constants for FeSn are highly anisotropic, with the nearest neighbor in-plane FM coupling constant J$_{in}$ $\approx$ -44 meV, and the nearest neighbor inter-plane AFM coupling constant J$_{int}$ $\approx$ 4.5 meV \cite{do_damped_2022}, there might be a second transition (FM-like transition) well above the Néel temperature. Upon cooling from the high-temperature paramagnetic phase, in-plane FM correlations might start to form first. The broad temperature range where the $c$-axis anomaly was observed in FeSn might suggest that magneto-elastic coupling arises from an increase in the in-plane spin order that begins tens of Kelvin above T$_{N}$. Similar magneto-elastic properties have been reported in the honeycomb lattice AFM material CrCl$_{3}$, where a negative thermal expansion in the $a$-axis on cooling from 50 K indicates FM coupling forming in-plane well above T$_{N}$ of 14 K \cite{schneeloch_gapless_2022}.

\begin{figure}[h!]
\begin{center}
\includegraphics[width=8.6cm]
{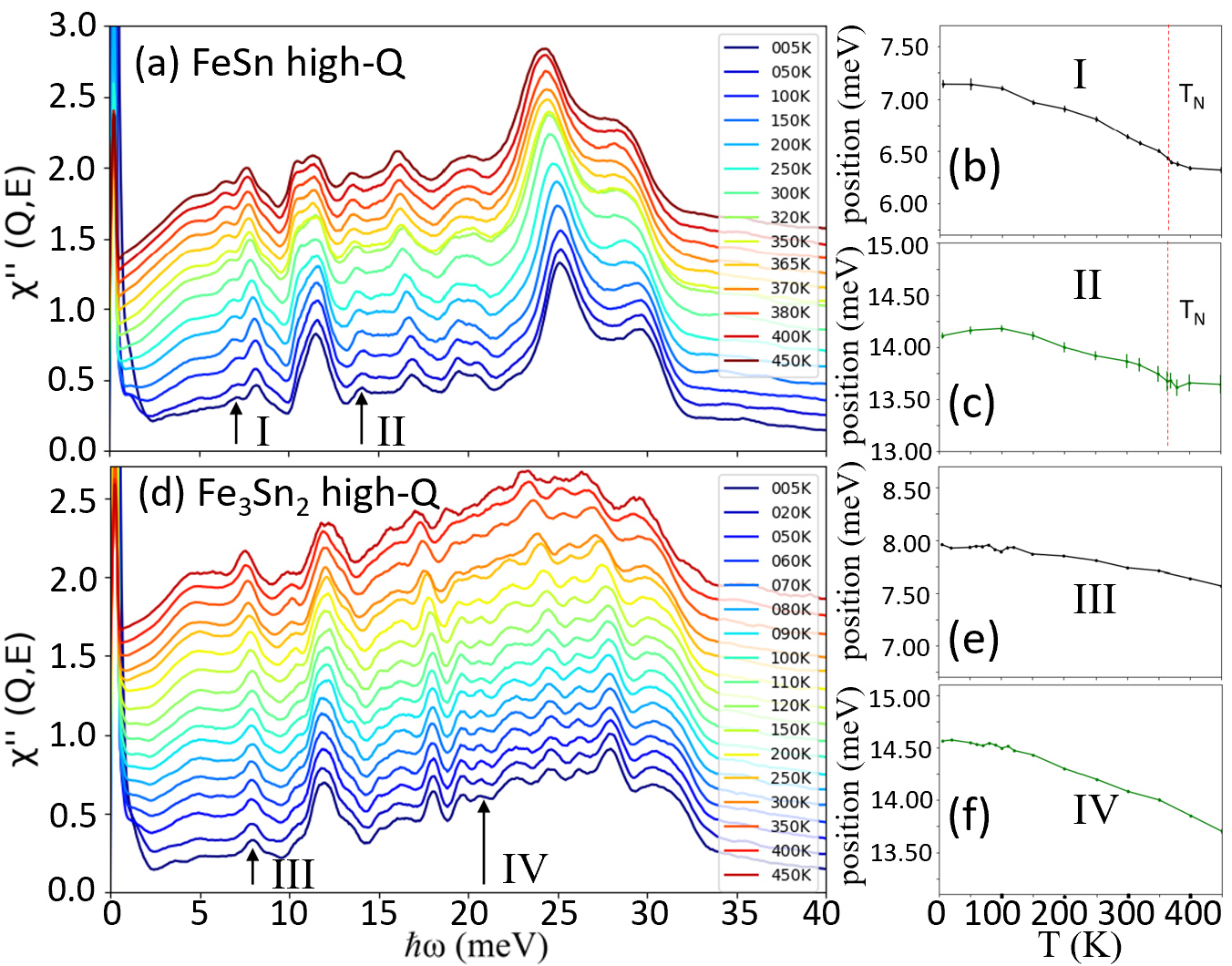}
\end{center}
\caption{(a) Dynamic susceptibility $\chi^{\prime \prime}(Q, E)$ for FeSn collected along the high-$Q$ trajectory as a function of temperature. (b,c) Energies of two FeSn phonon peaks (indicated with the black arrows in panel (a)) as a function of temperature, showing changes of the softening behavior across T$_{N}$. (d) Dynamic susceptibility $\chi^{\prime \prime}(Q, E)$ for Fe$_{3}$Sn$_{2}$ collected along the high-$Q$ trajectory as a function of temperature. (e,f) Energies of two Fe$_{3}$Sn$_{2}$ phonon peaks (indicated with the black arrows in panel (d)) as a function of temperature, showing usual softening.}
\label{fig:fig7}
\end{figure}

Shown in Fig.\ \ref{fig:fig7}(a) is the temperature dependence of the dynamic susceptibility (Bose-factor-corrected inelastic neutron scattering intensity), $\chi^{\prime \prime}(Q, E)$, for FeSn along the high-$Q$ trajectory. While the inelastic neutron scattering intensity along the low-$Q$ trajectory has approximately equal contributions from phonons and magnons, as shown earlier in Fig.\ \ref{fig:fig5}(g), the observed intensity along the high-$Q$ trajectory mainly comes from phonons due to the Q$^{2}$-dependence of the phonon intensity, with little contribution from the magnon, the intensity of which decreases with $Q$. Magnons contribute mostly at around 20 meV based on the simulations shown in Fig.\ \ref{fig:fig5}(d). At $\sim$12 meV, there is a clear change of the phonon peak intensity as a function of temperature. A shoulder develops at $\sim$10.5 meV upon warming, resulting in an overall change of the peak shape.

In addition, changes in the phonon softening behavior across the Néel temperature were observed for some phonon peaks in FeSn, suggesting spin-phonon coupling. Shown in Fig.\ \ref{fig:fig7}(b,c) are the energies of two selected phonon peaks (marked with the arrows in Fig.\ \ref{fig:fig7}(a)) as a function of temperature. Upon warming from low temperatures, both phonon peaks first show usual softening behavior, as phonon energy decreases with increasing temperature. However, when the temperature is above T$_{N}$, both phonon energies become constant.

For comparison, the temperature dependence of the dynamic susceptibility, $\chi^{\prime \prime}(Q, E)$, for Fe$_{3}$Sn$_{2}$ is shown in Fig.\ \ref{fig:fig7}(d). Little change in the phonon DOS was observed at all temperatures except for the peak at $\sim$22.5 meV, the intensity of which decreases slightly on warming from 250 to 300 K. However, no transition is known to occur in this temperature range. Note that for the measurements on Fe$_{3}$Sn$_{2}$, the strong peak at $\sim$12 meV contains instrument artifact, and whether or not phonon modes at this energy undergo similar changes as those observed in FeSn remains inconclusive.

For Fe$_{3}$Sn$_{2}$, some phonon measurements have been reported. Raman spectroscopy shows an anomaly in the linewidth for one of the A$_{g}$ modes at $\sim$100 K, which is thought due to the spin reorientation \cite{he_phonon_2022}. However, no signs of phonon anomaly was observed from the VISION data in this work. Shown in Fig.\ \ref{fig:fig7}(e,f) are the energies of two selected phonon peaks (marked with the arrows in Fig.\ \ref{fig:fig7}(d)) as a function of temperature. Usual softening behavior was observed across 100 K, suggesting little spin-phonon coupling in Fe$_{3}$Sn$_{2}$ up to at least 450 K.

\section{Conclusion}

In conclusion, the structural and dynamic properties of the magnetic topological quantum materials FeSn and Fe$_{3}$Sn$_{2}$ were investigated using neutron scattering and first-principles calculations. The AFM transition in FeSn is coupled with an anomaly in the $c$-axis lattice constant and a change in the phonon softening behavior, which are explained via DFT calculations as resulting from magneto-elastic and spin-phonon coupling. These features were not observed in Fe$_{3}$Sn$_{2}$, as measurements were performed at temperatures below T$_{C}$, indicating little change in the magnetism. Moreover, no evidence of electron localization was observed from the local structures of FeSn and Fe$_{3}$Sn$_{2}$. In addition, spin waves were observed in FeSn up to at least 450 K, suggesting a persistence of in-plane FM spin correlations well above T$_{N}$.

\section*{Acknowledgements}

This work has been supported by the Department of Energy, Grant number DE-FG02-01ER4592. A portion of this research used resources at the Spallation Neutron Source, a DOE Office of Science User Facility operated by Oak Ridge National Laboratory.

\bibliography{FeSn_reference}

\begin{thebibliography}{57}%
\makeatletter
\providecommand \@ifxundefined [1]{%
 \@ifx{#1\undefined}
}%
\providecommand \@ifnum [1]{%
 \ifnum #1\expandafter \@firstoftwo
 \else \expandafter \@secondoftwo
 \fi
}%
\providecommand \@ifx [1]{%
 \ifx #1\expandafter \@firstoftwo
 \else \expandafter \@secondoftwo
 \fi
}%
\providecommand \natexlab [1]{#1}%
\providecommand \enquote  [1]{``#1''}%
\providecommand \bibnamefont  [1]{#1}%
\providecommand \bibfnamefont [1]{#1}%
\providecommand \citenamefont [1]{#1}%
\providecommand \href@noop [0]{\@secondoftwo}%
\providecommand \href [0]{\begingroup \@sanitize@url \@href}%
\providecommand \@href[1]{\@@startlink{#1}\@@href}%
\providecommand \@@href[1]{\endgroup#1\@@endlink}%
\providecommand \@sanitize@url [0]{\catcode `\\12\catcode `\$12\catcode
  `\&12\catcode `\#12\catcode `\^12\catcode `\_12\catcode `\%12\relax}%
\providecommand \@@startlink[1]{}%
\providecommand \@@endlink[0]{}%
\providecommand \url  [0]{\begingroup\@sanitize@url \@url }%
\providecommand \@url [1]{\endgroup\@href {#1}{\urlprefix }}%
\providecommand \urlprefix  [0]{URL }%
\providecommand \Eprint [0]{\href }%
\providecommand \doibase [0]{http://dx.doi.org/}%
\providecommand \selectlanguage [0]{\@gobble}%
\providecommand \bibinfo  [0]{\@secondoftwo}%
\providecommand \bibfield  [0]{\@secondoftwo}%
\providecommand \translation [1]{[#1]}%
\providecommand \BibitemOpen [0]{}%
\providecommand \bibitemStop [0]{}%
\providecommand \bibitemNoStop [0]{.\EOS\space}%
\providecommand \EOS [0]{\spacefactor3000\relax}%
\providecommand \BibitemShut  [1]{\csname bibitem#1\endcsname}%
\let\auto@bib@innerbib\@empty
\bibitem [{\citenamefont {Tokura}\ \emph {et~al.}(2017)\citenamefont {Tokura},
  \citenamefont {Kawasaki},\ and\ \citenamefont
  {Nagaosa}}]{tokura_emergent_2017}%
  \BibitemOpen
  \bibfield  {author} {\bibinfo {author} {\bibfnamefont {Y.}~\bibnamefont
  {Tokura}}, \bibinfo {author} {\bibfnamefont {M.}~\bibnamefont {Kawasaki}}, \
  and\ \bibinfo {author} {\bibfnamefont {N.}~\bibnamefont {Nagaosa}},\
  }\bibfield  {title} {{\selectlanguage {english}\enquote {\bibinfo {title}
  {{Emergent} functions of quantum materials},}\ }}\href {\doibase
  10.1038/nphys4274} {\bibfield  {journal} {\bibinfo  {journal} {Nat. Phys.}\
  }\textbf {\bibinfo {volume} {13}},\ \bibinfo {pages} {1056--1068} (\bibinfo
  {year} {2017})},\ \bibinfo {note} {number: 11 Publisher: Nature Publishing
  Group}\BibitemShut {NoStop}%
\bibitem [{\citenamefont {Basov}\ \emph {et~al.}(2017)\citenamefont {Basov},
  \citenamefont {Averitt},\ and\ \citenamefont {Hsieh}}]{basov_towards_2017}%
  \BibitemOpen
  \bibfield  {author} {\bibinfo {author} {\bibfnamefont {D.~N.}\ \bibnamefont
  {Basov}}, \bibinfo {author} {\bibfnamefont {R.~D.}\ \bibnamefont {Averitt}},
  \ and\ \bibinfo {author} {\bibfnamefont {D.}~\bibnamefont {Hsieh}},\
  }\bibfield  {title} {{\selectlanguage {english}\enquote {\bibinfo {title}
  {{Towards} properties on demand in quantum materials},}\ }}\href {\doibase
  10.1038/nmat5017} {\bibfield  {journal} {\bibinfo  {journal} {Nat. Mater.}\
  }\textbf {\bibinfo {volume} {16}},\ \bibinfo {pages} {1077--1088} (\bibinfo
  {year} {2017})},\ \bibinfo {note} {number: 11 Publisher: Nature Publishing
  Group}\BibitemShut {NoStop}%
\bibitem [{\citenamefont {Emin}(1993)}]{emin_optical_1993}%
  \BibitemOpen
  \bibfield  {author} {\bibinfo {author} {\bibfnamefont {D.}~\bibnamefont
  {Emin}},\ }\bibfield  {title} {\enquote {\bibinfo {title} {{Optical}
  properties of large and small polarons and bipolarons},}\ }\href {\doibase
  10.1103/PhysRevB.48.13691} {\bibfield  {journal} {\bibinfo  {journal} {Phys.
  Rev. B}\ }\textbf {\bibinfo {volume} {48}},\ \bibinfo {pages} {13691--13702}
  (\bibinfo {year} {1993})},\ \bibinfo {note} {publisher: American Physical
  Society}\BibitemShut {NoStop}%
\bibitem [{\citenamefont {Stojchevska}\ \emph {et~al.}(2014)\citenamefont
  {Stojchevska}, \citenamefont {Vaskivskyi}, \citenamefont {Mertelj},
  \citenamefont {Kusar}, \citenamefont {Svetin}, \citenamefont {Brazovskii},\
  and\ \citenamefont {Mihailovic}}]{stojchevska_ultrafast_2014}%
  \BibitemOpen
  \bibfield  {author} {\bibinfo {author} {\bibfnamefont {L.}~\bibnamefont
  {Stojchevska}}, \bibinfo {author} {\bibfnamefont {I.}~\bibnamefont
  {Vaskivskyi}}, \bibinfo {author} {\bibfnamefont {T.}~\bibnamefont {Mertelj}},
  \bibinfo {author} {\bibfnamefont {P.}~\bibnamefont {Kusar}}, \bibinfo
  {author} {\bibfnamefont {D.}~\bibnamefont {Svetin}}, \bibinfo {author}
  {\bibfnamefont {S.}~\bibnamefont {Brazovskii}}, \ and\ \bibinfo {author}
  {\bibfnamefont {D.}~\bibnamefont {Mihailovic}},\ }\bibfield  {title}
  {\enquote {\bibinfo {title} {{Ultrafast} {Switching} to a {Stable} {Hidden}
  {Quantum} {State} in an {Electronic} {Crystal}},}\ }\href {\doibase
  10.1126/science.1241591} {\bibfield  {journal} {\bibinfo  {journal}
  {Science}\ }\textbf {\bibinfo {volume} {344}},\ \bibinfo {pages} {177--180}
  (\bibinfo {year} {2014})},\ \bibinfo {note} {publisher: American Association
  for the Advancement of Science}\BibitemShut {NoStop}%
\bibitem [{\citenamefont {Haug}\ \emph {et~al.}(2010)\citenamefont {Haug},
  \citenamefont {Hinkov}, \citenamefont {Sidis}, \citenamefont {Bourges},
  \citenamefont {Christensen}, \citenamefont {Ivanov}, \citenamefont {Keller},
  \citenamefont {Lin},\ and\ \citenamefont {Keimer}}]{haug_neutron_2010}%
  \BibitemOpen
  \bibfield  {author} {\bibinfo {author} {\bibfnamefont {D.}~\bibnamefont
  {Haug}}, \bibinfo {author} {\bibfnamefont {V.}~\bibnamefont {Hinkov}},
  \bibinfo {author} {\bibfnamefont {Y.}~\bibnamefont {Sidis}}, \bibinfo
  {author} {\bibfnamefont {P.}~\bibnamefont {Bourges}}, \bibinfo {author}
  {\bibfnamefont {N.~B.}\ \bibnamefont {Christensen}}, \bibinfo {author}
  {\bibfnamefont {A.}~\bibnamefont {Ivanov}}, \bibinfo {author} {\bibfnamefont
  {T.}~\bibnamefont {Keller}}, \bibinfo {author} {\bibfnamefont {C.~T.}\
  \bibnamefont {Lin}}, \ and\ \bibinfo {author} {\bibfnamefont
  {B.}~\bibnamefont {Keimer}},\ }\bibfield  {title} {{\selectlanguage
  {english}\enquote {\bibinfo {title} {{Neutron} scattering study of the
  magnetic phase diagram of underdoped {YBa$_{2}$Cu$_{3}$O$_{6+x}$}},}\ }}\href
  {\doibase 10.1088/1367-2630/12/10/105006} {\bibfield  {journal} {\bibinfo
  {journal} {New J. Phys.}\ }\textbf {\bibinfo {volume} {12}},\ \bibinfo
  {pages} {105006} (\bibinfo {year} {2010})}\BibitemShut {NoStop}%
\bibitem [{\citenamefont {Yan}\ and\ \citenamefont
  {Felser}(2017)}]{yan_topological_2017}%
  \BibitemOpen
  \bibfield  {author} {\bibinfo {author} {\bibfnamefont {B.}~\bibnamefont
  {Yan}}\ and\ \bibinfo {author} {\bibfnamefont {C.}~\bibnamefont {Felser}},\
  }\bibfield  {title} {\enquote {\bibinfo {title} {{Topological} {Materials}:
  {Weyl} {Semimetals}},}\ }\href {\doibase
  10.1146/annurev-conmatphys-031016-025458} {\bibfield  {journal} {\bibinfo
  {journal} {Annu. Rev. Condens. Matter Phys.}\ }\textbf {\bibinfo {volume}
  {8}},\ \bibinfo {pages} {337--354} (\bibinfo {year} {2017})},\ \bibinfo
  {note} {\_eprint:
  https://doi.org/10.1146/annurev-conmatphys-031016-025458}\BibitemShut
  {NoStop}%
\bibitem [{\citenamefont {Qi}\ and\ \citenamefont
  {Zhang}(2011)}]{qi_topological_2011}%
  \BibitemOpen
  \bibfield  {author} {\bibinfo {author} {\bibfnamefont {X.~L.}\ \bibnamefont
  {Qi}}\ and\ \bibinfo {author} {\bibfnamefont {S.~C.}\ \bibnamefont {Zhang}},\
  }\bibfield  {title} {\enquote {\bibinfo {title} {{Topological} insulators and
  superconductors},}\ }\href {\doibase 10.1103/RevModPhys.83.1057} {\bibfield
  {journal} {\bibinfo  {journal} {Rev. Mod. Phys.}\ }\textbf {\bibinfo {volume}
  {83}},\ \bibinfo {pages} {1057--1110} (\bibinfo {year} {2011})},\ \bibinfo
  {note} {publisher: American Physical Society}\BibitemShut {NoStop}%
\bibitem [{\citenamefont {Hasan}\ and\ \citenamefont
  {Kane}(2010)}]{hasan_colloquium_2010}%
  \BibitemOpen
  \bibfield  {author} {\bibinfo {author} {\bibfnamefont {M.~Z.}\ \bibnamefont
  {Hasan}}\ and\ \bibinfo {author} {\bibfnamefont {C.~L.}\ \bibnamefont
  {Kane}},\ }\bibfield  {title} {\enquote {\bibinfo {title} {{Colloquium}:
  {Topological} insulators},}\ }\href {\doibase 10.1103/RevModPhys.82.3045}
  {\bibfield  {journal} {\bibinfo  {journal} {Rev. Mod. Phys.}\ }\textbf
  {\bibinfo {volume} {82}},\ \bibinfo {pages} {3045--3067} (\bibinfo {year}
  {2010})},\ \bibinfo {note} {publisher: American Physical Society}\BibitemShut
  {NoStop}%
\bibitem [{\citenamefont {Aoki}\ \emph {et~al.}(1996)\citenamefont {Aoki},
  \citenamefont {Ando},\ and\ \citenamefont
  {Matsumura}}]{aoki_hofstadter_1996}%
  \BibitemOpen
  \bibfield  {author} {\bibinfo {author} {\bibfnamefont {H.}~\bibnamefont
  {Aoki}}, \bibinfo {author} {\bibfnamefont {M.}~\bibnamefont {Ando}}, \ and\
  \bibinfo {author} {\bibfnamefont {H.}~\bibnamefont {Matsumura}},\ }\bibfield
  {title} {\enquote {\bibinfo {title} {{Hofstadter} butterflies for flat
  bands},}\ }\href {\doibase 10.1103/PhysRevB.54.R17296} {\bibfield  {journal}
  {\bibinfo  {journal} {Phys. Rev. B}\ }\textbf {\bibinfo {volume} {54}},\
  \bibinfo {pages} {R17296--R17299} (\bibinfo {year} {1996})},\ \bibinfo {note}
  {publisher: American Physical Society}\BibitemShut {NoStop}%
\bibitem [{\citenamefont {Deng}\ \emph {et~al.}(2003)\citenamefont {Deng},
  \citenamefont {Simon},\ and\ \citenamefont {Köhler}}]{deng_origin_2003}%
  \BibitemOpen
  \bibfield  {author} {\bibinfo {author} {\bibfnamefont {S.}~\bibnamefont
  {Deng}}, \bibinfo {author} {\bibfnamefont {A.}~\bibnamefont {Simon}}, \ and\
  \bibinfo {author} {\bibfnamefont {J.}~\bibnamefont {Köhler}},\ }\bibfield
  {title} {{\selectlanguage {english}\enquote {\bibinfo {title} {{The} origin
  of a flat band},}\ }}\href {\doibase 10.1016/S0022-4596(03)00239-1}
  {\bibfield  {journal} {\bibinfo  {journal} {J. Solid State Chem.}\ }\bibinfo
  {series} {Special issue on {The} {Impact} of {Theoretical} {Methods} on
  {Solid}-{State} {Chemistry}},\ \textbf {\bibinfo {volume} {176}},\ \bibinfo
  {pages} {412--416} (\bibinfo {year} {2003})}\BibitemShut {NoStop}%
\bibitem [{\citenamefont {Wu}\ \emph {et~al.}(2007)\citenamefont {Wu},
  \citenamefont {Bergman}, \citenamefont {Balents},\ and\ \citenamefont
  {Sarma}}]{wu_flat_2007}%
  \BibitemOpen
  \bibfield  {author} {\bibinfo {author} {\bibfnamefont {C.}~\bibnamefont
  {Wu}}, \bibinfo {author} {\bibfnamefont {D.}~\bibnamefont {Bergman}},
  \bibinfo {author} {\bibfnamefont {L.}~\bibnamefont {Balents}}, \ and\
  \bibinfo {author} {\bibfnamefont {S.~Das}\ \bibnamefont {Sarma}},\ }\bibfield
   {title} {\enquote {\bibinfo {title} {{Flat} {Bands} and {Wigner}
  {Crystallization} in the {Honeycomb} {Optical} {Lattice}},}\ }\href {\doibase
  10.1103/PhysRevLett.99.070401} {\bibfield  {journal} {\bibinfo  {journal}
  {Phys. Rev. Lett.}\ }\textbf {\bibinfo {volume} {99}},\ \bibinfo {pages}
  {070401} (\bibinfo {year} {2007})},\ \bibinfo {note} {publisher: American
  Physical Society}\BibitemShut {NoStop}%
\bibitem [{\citenamefont {Jacqmin}\ \emph {et~al.}(2014)\citenamefont
  {Jacqmin}, \citenamefont {Carusotto}, \citenamefont {Sagnes}, \citenamefont
  {Abbarchi}, \citenamefont {Solnyshkov}, \citenamefont {Malpuech},
  \citenamefont {Galopin}, \citenamefont {Lemaître}, \citenamefont {Bloch},\
  and\ \citenamefont {Amo}}]{jacqmin_direct_2014}%
  \BibitemOpen
  \bibfield  {author} {\bibinfo {author} {\bibfnamefont {T.}~\bibnamefont
  {Jacqmin}}, \bibinfo {author} {\bibfnamefont {I.}~\bibnamefont {Carusotto}},
  \bibinfo {author} {\bibfnamefont {I.}~\bibnamefont {Sagnes}}, \bibinfo
  {author} {\bibfnamefont {M.}~\bibnamefont {Abbarchi}}, \bibinfo {author}
  {\bibfnamefont {D.~D.}\ \bibnamefont {Solnyshkov}}, \bibinfo {author}
  {\bibfnamefont {G.}~\bibnamefont {Malpuech}}, \bibinfo {author}
  {\bibfnamefont {E.}~\bibnamefont {Galopin}}, \bibinfo {author} {\bibfnamefont
  {A.}~\bibnamefont {Lemaître}}, \bibinfo {author} {\bibfnamefont
  {J.}~\bibnamefont {Bloch}}, \ and\ \bibinfo {author} {\bibfnamefont
  {A.}~\bibnamefont {Amo}},\ }\bibfield  {title} {\enquote {\bibinfo {title}
  {{Direct} {Observation} of {Dirac} {Cones} and a {Flatband} in a {Honeycomb}
  {Lattice} for {Polaritons}},}\ }\href {\doibase
  10.1103/PhysRevLett.112.116402} {\bibfield  {journal} {\bibinfo  {journal}
  {Phys. Rev. Lett.}\ }\textbf {\bibinfo {volume} {112}},\ \bibinfo {pages}
  {116402} (\bibinfo {year} {2014})},\ \bibinfo {note} {publisher: American
  Physical Society}\BibitemShut {NoStop}%
\bibitem [{\citenamefont {Mukherjee}\ \emph {et~al.}(2015)\citenamefont
  {Mukherjee}, \citenamefont {Spracklen}, \citenamefont {Choudhury},
  \citenamefont {Goldman}, \citenamefont {Öhberg}, \citenamefont {Andersson},\
  and\ \citenamefont {Thomson}}]{mukherjee_observation_2015}%
  \BibitemOpen
  \bibfield  {author} {\bibinfo {author} {\bibfnamefont {S.}~\bibnamefont
  {Mukherjee}}, \bibinfo {author} {\bibfnamefont {A.}~\bibnamefont
  {Spracklen}}, \bibinfo {author} {\bibfnamefont {D.}~\bibnamefont
  {Choudhury}}, \bibinfo {author} {\bibfnamefont {N.}~\bibnamefont {Goldman}},
  \bibinfo {author} {\bibfnamefont {P.}~\bibnamefont {Öhberg}}, \bibinfo
  {author} {\bibfnamefont {E.}~\bibnamefont {Andersson}}, \ and\ \bibinfo
  {author} {\bibfnamefont {R.~R.}\ \bibnamefont {Thomson}},\ }\bibfield
  {title} {\enquote {\bibinfo {title} {{Observation} of a {Localized}
  {Flat}-{Band} {State} in a {Photonic} {Lieb} {Lattice}},}\ }\href {\doibase
  10.1103/PhysRevLett.114.245504} {\bibfield  {journal} {\bibinfo  {journal}
  {Phys. Rev. Lett.}\ }\textbf {\bibinfo {volume} {114}},\ \bibinfo {pages}
  {245504} (\bibinfo {year} {2015})},\ \bibinfo {note} {publisher: American
  Physical Society}\BibitemShut {NoStop}%
\bibitem [{\citenamefont {Tilak}\ \emph {et~al.}(2021)\citenamefont {Tilak},
  \citenamefont {Lai}, \citenamefont {Wu}, \citenamefont {Zhang}, \citenamefont
  {Xu}, \citenamefont {Ribeiro}, \citenamefont {Canfield},\ and\ \citenamefont
  {Andrei}}]{tilak_flat_2021}%
  \BibitemOpen
  \bibfield  {author} {\bibinfo {author} {\bibfnamefont {N.}~\bibnamefont
  {Tilak}}, \bibinfo {author} {\bibfnamefont {X.}~\bibnamefont {Lai}}, \bibinfo
  {author} {\bibfnamefont {S.}~\bibnamefont {Wu}}, \bibinfo {author}
  {\bibfnamefont {Z.}~\bibnamefont {Zhang}}, \bibinfo {author} {\bibfnamefont
  {M.}~\bibnamefont {Xu}}, \bibinfo {author} {\bibfnamefont {R.~D.~A.}\
  \bibnamefont {Ribeiro}}, \bibinfo {author} {\bibfnamefont {P.~C.}\
  \bibnamefont {Canfield}}, \ and\ \bibinfo {author} {\bibfnamefont {E.~Y.}\
  \bibnamefont {Andrei}},\ }\bibfield  {title} {{\selectlanguage
  {english}\enquote {\bibinfo {title} {{Flat} band carrier confinement in
  magic-angle twisted bilayer graphene},}\ }}\href {\doibase
  10.1038/s41467-021-24480-3} {\bibfield  {journal} {\bibinfo  {journal} {Nat.
  Commun.}\ }\textbf {\bibinfo {volume} {12}},\ \bibinfo {pages} {4180}
  (\bibinfo {year} {2021})},\ \bibinfo {note} {number: 1 Publisher: Nature
  Publishing Group}\BibitemShut {NoStop}%
\bibitem [{\citenamefont {Kang}\ \emph
  {et~al.}(2020{\natexlab{a}})\citenamefont {Kang}, \citenamefont {Fang},
  \citenamefont {Ye}, \citenamefont {Po}, \citenamefont {Denlinger},
  \citenamefont {Jozwiak}, \citenamefont {Bostwick}, \citenamefont {Rotenberg},
  \citenamefont {Kaxiras}, \citenamefont {Checkelsky},\ and\ \citenamefont
  {Comin}}]{kang_topological_2020}%
  \BibitemOpen
  \bibfield  {author} {\bibinfo {author} {\bibfnamefont {M.}~\bibnamefont
  {Kang}}, \bibinfo {author} {\bibfnamefont {S.}~\bibnamefont {Fang}}, \bibinfo
  {author} {\bibfnamefont {L.}~\bibnamefont {Ye}}, \bibinfo {author}
  {\bibfnamefont {H.~C.}\ \bibnamefont {Po}}, \bibinfo {author} {\bibfnamefont
  {J.}~\bibnamefont {Denlinger}}, \bibinfo {author} {\bibfnamefont
  {C.}~\bibnamefont {Jozwiak}}, \bibinfo {author} {\bibfnamefont
  {A.}~\bibnamefont {Bostwick}}, \bibinfo {author} {\bibfnamefont
  {E.}~\bibnamefont {Rotenberg}}, \bibinfo {author} {\bibfnamefont
  {E.}~\bibnamefont {Kaxiras}}, \bibinfo {author} {\bibfnamefont {J.~G.}\
  \bibnamefont {Checkelsky}}, \ and\ \bibinfo {author} {\bibfnamefont
  {R.}~\bibnamefont {Comin}},\ }\bibfield  {title} {{\selectlanguage
  {english}\enquote {\bibinfo {title} {{Topological} flat bands in frustrated
  kagome lattice {CoSn}},}\ }}\href {\doibase 10.1038/s41467-020-17465-1}
  {\bibfield  {journal} {\bibinfo  {journal} {Nat. Commun.}\ }\textbf {\bibinfo
  {volume} {11}},\ \bibinfo {pages} {4004} (\bibinfo {year}
  {2020}{\natexlab{a}})},\ \bibinfo {note} {number: 1 Publisher: Nature
  Publishing Group}\BibitemShut {NoStop}%
\bibitem [{\citenamefont {Lin}\ \emph {et~al.}(2018)\citenamefont {Lin},
  \citenamefont {Choi}, \citenamefont {Zhang}, \citenamefont {Qin},
  \citenamefont {Yi}, \citenamefont {Wang}, \citenamefont {Li}, \citenamefont
  {Wang}, \citenamefont {Zhang}, \citenamefont {Sun}, \citenamefont {Wei},
  \citenamefont {Zhang}, \citenamefont {Guo}, \citenamefont {Lu}, \citenamefont
  {Cho}, \citenamefont {Zeng},\ and\ \citenamefont
  {Zhang}}]{lin_flatbands_2018}%
  \BibitemOpen
  \bibfield  {author} {\bibinfo {author} {\bibfnamefont {Z.}~\bibnamefont
  {Lin}}, \bibinfo {author} {\bibfnamefont {J.~H.}\ \bibnamefont {Choi}},
  \bibinfo {author} {\bibfnamefont {Q.}~\bibnamefont {Zhang}}, \bibinfo
  {author} {\bibfnamefont {W.}~\bibnamefont {Qin}}, \bibinfo {author}
  {\bibfnamefont {S.}~\bibnamefont {Yi}}, \bibinfo {author} {\bibfnamefont
  {P.}~\bibnamefont {Wang}}, \bibinfo {author} {\bibfnamefont {L.}~\bibnamefont
  {Li}}, \bibinfo {author} {\bibfnamefont {Y.}~\bibnamefont {Wang}}, \bibinfo
  {author} {\bibfnamefont {H.}~\bibnamefont {Zhang}}, \bibinfo {author}
  {\bibfnamefont {Z.}~\bibnamefont {Sun}}, \bibinfo {author} {\bibfnamefont
  {L.}~\bibnamefont {Wei}}, \bibinfo {author} {\bibfnamefont {S.}~\bibnamefont
  {Zhang}}, \bibinfo {author} {\bibfnamefont {T.}~\bibnamefont {Guo}}, \bibinfo
  {author} {\bibfnamefont {Q.}~\bibnamefont {Lu}}, \bibinfo {author}
  {\bibfnamefont {J.~H.}\ \bibnamefont {Cho}}, \bibinfo {author} {\bibfnamefont
  {C.}~\bibnamefont {Zeng}}, \ and\ \bibinfo {author} {\bibfnamefont
  {Z.}~\bibnamefont {Zhang}},\ }\bibfield  {title} {\enquote {\bibinfo {title}
  {{Flatbands} and {Emergent} {Ferromagnetic} {Ordering} in {Fe$_{3}$Sn$_{2}$}
  {Kagome} {Lattices}},}\ }\href {\doibase 10.1103/PhysRevLett.121.096401}
  {\bibfield  {journal} {\bibinfo  {journal} {Phys. Rev. Lett.}\ }\textbf
  {\bibinfo {volume} {121}},\ \bibinfo {pages} {096401} (\bibinfo {year}
  {2018})},\ \bibinfo {note} {publisher: American Physical Society}\BibitemShut
  {NoStop}%
\bibitem [{\citenamefont {Ito}\ \emph {et~al.}(1999)\citenamefont {Ito},
  \citenamefont {Kumigashira}, \citenamefont {Kim}, \citenamefont {Takahashi},
  \citenamefont {Kimura}, \citenamefont {Haga}, \citenamefont {Yamamoto},
  \citenamefont {Ōnuki},\ and\ \citenamefont
  {Harima}}]{ito_high-resolution_1999}%
  \BibitemOpen
  \bibfield  {author} {\bibinfo {author} {\bibfnamefont {T.}~\bibnamefont
  {Ito}}, \bibinfo {author} {\bibfnamefont {H.}~\bibnamefont {Kumigashira}},
  \bibinfo {author} {\bibfnamefont {H.~D.}\ \bibnamefont {Kim}}, \bibinfo
  {author} {\bibfnamefont {T.}~\bibnamefont {Takahashi}}, \bibinfo {author}
  {\bibfnamefont {N.}~\bibnamefont {Kimura}}, \bibinfo {author} {\bibfnamefont
  {Y.}~\bibnamefont {Haga}}, \bibinfo {author} {\bibfnamefont {E.}~\bibnamefont
  {Yamamoto}}, \bibinfo {author} {\bibfnamefont {Y.}~\bibnamefont {Ōnuki}}, \
  and\ \bibinfo {author} {\bibfnamefont {H.}~\bibnamefont {Harima}},\
  }\bibfield  {title} {\enquote {\bibinfo {title} {{High}-resolution
  angle-resolved photoemission study of the heavy-fermion superconductor
  {UPt$_{3}$}},}\ }\href {\doibase 10.1103/PhysRevB.59.8923} {\bibfield
  {journal} {\bibinfo  {journal} {Phys. Rev. B}\ }\textbf {\bibinfo {volume}
  {59}},\ \bibinfo {pages} {8923--8929} (\bibinfo {year} {1999})},\ \bibinfo
  {note} {publisher: American Physical Society}\BibitemShut {NoStop}%
\bibitem [{\citenamefont {Kang}\ \emph
  {et~al.}(2020{\natexlab{b}})\citenamefont {Kang}, \citenamefont {Ye},
  \citenamefont {Fang}, \citenamefont {You}, \citenamefont {Levitan},
  \citenamefont {Han}, \citenamefont {Facio}, \citenamefont {Jozwiak},
  \citenamefont {Bostwick}, \citenamefont {Rotenberg}, \citenamefont {Chan},
  \citenamefont {McDonald}, \citenamefont {Graf}, \citenamefont {Kaznatcheev},
  \citenamefont {Vescovo}, \citenamefont {Bell}, \citenamefont {Kaxiras},
  \citenamefont {van~den Brink}, \citenamefont {Richter}, \citenamefont
  {Ghimire}, \citenamefont {Checkelsky},\ and\ \citenamefont
  {Comin}}]{kang_dirac_2020}%
  \BibitemOpen
  \bibfield  {author} {\bibinfo {author} {\bibfnamefont {M.}~\bibnamefont
  {Kang}}, \bibinfo {author} {\bibfnamefont {L.}~\bibnamefont {Ye}}, \bibinfo
  {author} {\bibfnamefont {S.}~\bibnamefont {Fang}}, \bibinfo {author}
  {\bibfnamefont {J.~S.}\ \bibnamefont {You}}, \bibinfo {author} {\bibfnamefont
  {A.}~\bibnamefont {Levitan}}, \bibinfo {author} {\bibfnamefont
  {M.}~\bibnamefont {Han}}, \bibinfo {author} {\bibfnamefont {J.~I.}\
  \bibnamefont {Facio}}, \bibinfo {author} {\bibfnamefont {C.}~\bibnamefont
  {Jozwiak}}, \bibinfo {author} {\bibfnamefont {A.}~\bibnamefont {Bostwick}},
  \bibinfo {author} {\bibfnamefont {E.}~\bibnamefont {Rotenberg}}, \bibinfo
  {author} {\bibfnamefont {M.~K.}\ \bibnamefont {Chan}}, \bibinfo {author}
  {\bibfnamefont {R.~D.}\ \bibnamefont {McDonald}}, \bibinfo {author}
  {\bibfnamefont {D.}~\bibnamefont {Graf}}, \bibinfo {author} {\bibfnamefont
  {K.}~\bibnamefont {Kaznatcheev}}, \bibinfo {author} {\bibfnamefont
  {E.}~\bibnamefont {Vescovo}}, \bibinfo {author} {\bibfnamefont {D.~C.}\
  \bibnamefont {Bell}}, \bibinfo {author} {\bibfnamefont {E.}~\bibnamefont
  {Kaxiras}}, \bibinfo {author} {\bibfnamefont {J.}~\bibnamefont {van~den
  Brink}}, \bibinfo {author} {\bibfnamefont {M.}~\bibnamefont {Richter}},
  \bibinfo {author} {\bibfnamefont {M.~P.}\ \bibnamefont {Ghimire}}, \bibinfo
  {author} {\bibfnamefont {J.~G.}\ \bibnamefont {Checkelsky}}, \ and\ \bibinfo
  {author} {\bibfnamefont {R.}~\bibnamefont {Comin}},\ }\bibfield  {title}
  {{\selectlanguage {english}\enquote {\bibinfo {title} {{Dirac} fermions and
  flat bands in the ideal kagome metal {FeSn}},}\ }}\href {\doibase
  10.1038/s41563-019-0531-0} {\bibfield  {journal} {\bibinfo  {journal} {Nat.
  Mater.}\ }\textbf {\bibinfo {volume} {19}},\ \bibinfo {pages} {163--169}
  (\bibinfo {year} {2020}{\natexlab{b}})},\ \bibinfo {note} {number: 2
  Publisher: Nature Publishing Group}\BibitemShut {NoStop}%
\bibitem [{\citenamefont {Li}\ \emph {et~al.}(2018)\citenamefont {Li},
  \citenamefont {Zhuang}, \citenamefont {Wang}, \citenamefont {Feng},
  \citenamefont {Gao}, \citenamefont {Xu}, \citenamefont {Hao}, \citenamefont
  {Wang}, \citenamefont {Zhang}, \citenamefont {Wu}, \citenamefont {Dou},
  \citenamefont {Chen}, \citenamefont {Hu},\ and\ \citenamefont
  {Du}}]{li_realization_2018}%
  \BibitemOpen
  \bibfield  {author} {\bibinfo {author} {\bibfnamefont {Z.}~\bibnamefont
  {Li}}, \bibinfo {author} {\bibfnamefont {J.}~\bibnamefont {Zhuang}}, \bibinfo
  {author} {\bibfnamefont {L.}~\bibnamefont {Wang}}, \bibinfo {author}
  {\bibfnamefont {H.}~\bibnamefont {Feng}}, \bibinfo {author} {\bibfnamefont
  {Q.}~\bibnamefont {Gao}}, \bibinfo {author} {\bibfnamefont {X.}~\bibnamefont
  {Xu}}, \bibinfo {author} {\bibfnamefont {W.}~\bibnamefont {Hao}}, \bibinfo
  {author} {\bibfnamefont {X.}~\bibnamefont {Wang}}, \bibinfo {author}
  {\bibfnamefont {C.}~\bibnamefont {Zhang}}, \bibinfo {author} {\bibfnamefont
  {K.}~\bibnamefont {Wu}}, \bibinfo {author} {\bibfnamefont {S.~X.}\
  \bibnamefont {Dou}}, \bibinfo {author} {\bibfnamefont {L.}~\bibnamefont
  {Chen}}, \bibinfo {author} {\bibfnamefont {Z.}~\bibnamefont {Hu}}, \ and\
  \bibinfo {author} {\bibfnamefont {Y.}~\bibnamefont {Du}},\ }\bibfield
  {title} {\enquote {\bibinfo {title} {{Realization} of flat band with possible
  nontrivial topology in electronic {Kagome} lattice},}\ }\href {\doibase
  10.1126/sciadv.aau4511} {\bibfield  {journal} {\bibinfo  {journal} {Sci.
  Adv.}\ }\textbf {\bibinfo {volume} {4}},\ \bibinfo {pages} {eaau4511}
  (\bibinfo {year} {2018})},\ \bibinfo {note} {publisher: American Association
  for the Advancement of Science}\BibitemShut {NoStop}%
\bibitem [{\citenamefont {Xu}\ \emph {et~al.}(2020)\citenamefont {Xu},
  \citenamefont {Zhao}, \citenamefont {Yi}, \citenamefont {Wang}, \citenamefont
  {Yin}, \citenamefont {Wang}, \citenamefont {Hu}, \citenamefont {Wang},
  \citenamefont {Liu}, \citenamefont {Xu}, \citenamefont {Lu}, \citenamefont
  {Soluyanov}, \citenamefont {Lei}, \citenamefont {Shi}, \citenamefont {Luo},\
  and\ \citenamefont {Chen}}]{xu_electronic_2020}%
  \BibitemOpen
  \bibfield  {author} {\bibinfo {author} {\bibfnamefont {Y.}~\bibnamefont
  {Xu}}, \bibinfo {author} {\bibfnamefont {J.}~\bibnamefont {Zhao}}, \bibinfo
  {author} {\bibfnamefont {C.}~\bibnamefont {Yi}}, \bibinfo {author}
  {\bibfnamefont {Q.}~\bibnamefont {Wang}}, \bibinfo {author} {\bibfnamefont
  {Q.}~\bibnamefont {Yin}}, \bibinfo {author} {\bibfnamefont {Y.}~\bibnamefont
  {Wang}}, \bibinfo {author} {\bibfnamefont {X.}~\bibnamefont {Hu}}, \bibinfo
  {author} {\bibfnamefont {L.}~\bibnamefont {Wang}}, \bibinfo {author}
  {\bibfnamefont {E.}~\bibnamefont {Liu}}, \bibinfo {author} {\bibfnamefont
  {G.}~\bibnamefont {Xu}}, \bibinfo {author} {\bibfnamefont {L.}~\bibnamefont
  {Lu}}, \bibinfo {author} {\bibfnamefont {A.~A.}\ \bibnamefont {Soluyanov}},
  \bibinfo {author} {\bibfnamefont {H.}~\bibnamefont {Lei}}, \bibinfo {author}
  {\bibfnamefont {Y.}~\bibnamefont {Shi}}, \bibinfo {author} {\bibfnamefont
  {J.}~\bibnamefont {Luo}}, \ and\ \bibinfo {author} {\bibfnamefont {Z.~G.}\
  \bibnamefont {Chen}},\ }\bibfield  {title} {{\selectlanguage
  {english}\enquote {\bibinfo {title} {Electronic correlations and flattened
  band in magnetic {Weyl} semimetal candidate {Co$_{3}$Sn$_{2}$S$_{2}$}},}\
  }}\href {\doibase 10.1038/s41467-020-17234-0} {\bibfield  {journal} {\bibinfo
   {journal} {Nat. Commun.}\ }\textbf {\bibinfo {volume} {11}},\ \bibinfo
  {pages} {3985} (\bibinfo {year} {2020})},\ \bibinfo {note} {number: 1
  Publisher: Nature Publishing Group}\BibitemShut {NoStop}%
\bibitem [{\citenamefont {Ye}\ \emph {et~al.}(2018)\citenamefont {Ye},
  \citenamefont {Kang}, \citenamefont {Liu}, \citenamefont {von Cube},
  \citenamefont {Wicker}, \citenamefont {Suzuki}, \citenamefont {Jozwiak},
  \citenamefont {Bostwick}, \citenamefont {Rotenberg}, \citenamefont {Bell},
  \citenamefont {Fu}, \citenamefont {Comin},\ and\ \citenamefont
  {Checkelsky}}]{ye_massive_2018}%
  \BibitemOpen
  \bibfield  {author} {\bibinfo {author} {\bibfnamefont {L.}~\bibnamefont
  {Ye}}, \bibinfo {author} {\bibfnamefont {M.}~\bibnamefont {Kang}}, \bibinfo
  {author} {\bibfnamefont {J.}~\bibnamefont {Liu}}, \bibinfo {author}
  {\bibfnamefont {F.}~\bibnamefont {von Cube}}, \bibinfo {author}
  {\bibfnamefont {C.~R.}\ \bibnamefont {Wicker}}, \bibinfo {author}
  {\bibfnamefont {T.}~\bibnamefont {Suzuki}}, \bibinfo {author} {\bibfnamefont
  {C.}~\bibnamefont {Jozwiak}}, \bibinfo {author} {\bibfnamefont
  {A.}~\bibnamefont {Bostwick}}, \bibinfo {author} {\bibfnamefont
  {E.}~\bibnamefont {Rotenberg}}, \bibinfo {author} {\bibfnamefont {D.~C.}\
  \bibnamefont {Bell}}, \bibinfo {author} {\bibfnamefont {L.}~\bibnamefont
  {Fu}}, \bibinfo {author} {\bibfnamefont {R.}~\bibnamefont {Comin}}, \ and\
  \bibinfo {author} {\bibfnamefont {J.~G.}\ \bibnamefont {Checkelsky}},\
  }\bibfield  {title} {{\selectlanguage {english}\enquote {\bibinfo {title}
  {Massive {Dirac} fermions in a ferromagnetic kagome metal},}\ }}\href
  {\doibase 10.1038/nature25987} {\bibfield  {journal} {\bibinfo  {journal}
  {Nature}\ }\textbf {\bibinfo {volume} {555}},\ \bibinfo {pages} {638--642}
  (\bibinfo {year} {2018})},\ \bibinfo {note} {number: 7698 Publisher: Nature
  Publishing Group}\BibitemShut {NoStop}%
\bibitem [{\citenamefont {Li}\ \emph {et~al.}(2021{\natexlab{a}})\citenamefont
  {Li}, \citenamefont {Wang}, \citenamefont {Wang}, \citenamefont {Yuan},
  \citenamefont {Song}, \citenamefont {Lou}, \citenamefont {Liu}, \citenamefont
  {Huang}, \citenamefont {Liu}, \citenamefont {Lei}, \citenamefont {Yin},\ and\
  \citenamefont {Wang}}]{li_dirac_2021}%
  \BibitemOpen
  \bibfield  {author} {\bibinfo {author} {\bibfnamefont {M.}~\bibnamefont
  {Li}}, \bibinfo {author} {\bibfnamefont {Q.}~\bibnamefont {Wang}}, \bibinfo
  {author} {\bibfnamefont {G.}~\bibnamefont {Wang}}, \bibinfo {author}
  {\bibfnamefont {Z.}~\bibnamefont {Yuan}}, \bibinfo {author} {\bibfnamefont
  {W.}~\bibnamefont {Song}}, \bibinfo {author} {\bibfnamefont {R.}~\bibnamefont
  {Lou}}, \bibinfo {author} {\bibfnamefont {Z.}~\bibnamefont {Liu}}, \bibinfo
  {author} {\bibfnamefont {Y.}~\bibnamefont {Huang}}, \bibinfo {author}
  {\bibfnamefont {Z.}~\bibnamefont {Liu}}, \bibinfo {author} {\bibfnamefont
  {H.}~\bibnamefont {Lei}}, \bibinfo {author} {\bibfnamefont {Z.}~\bibnamefont
  {Yin}}, \ and\ \bibinfo {author} {\bibfnamefont {S.}~\bibnamefont {Wang}},\
  }\bibfield  {title} {{\selectlanguage {english}\enquote {\bibinfo {title}
  {{Dirac} cone, flat band and saddle point in kagome magnet
  {YMn$_{6}$Sn$_{6}$}},}\ }}\href {\doibase 10.1038/s41467-021-23536-8}
  {\bibfield  {journal} {\bibinfo  {journal} {Nat. Commun.}\ }\textbf {\bibinfo
  {volume} {12}},\ \bibinfo {pages} {3129} (\bibinfo {year}
  {2021}{\natexlab{a}})},\ \bibinfo {note} {number: 1 Publisher: Nature
  Publishing Group}\BibitemShut {NoStop}%
\bibitem [{\citenamefont {Han}\ \emph {et~al.}(2012)\citenamefont {Han},
  \citenamefont {Helton}, \citenamefont {Chu}, \citenamefont {Nocera},
  \citenamefont {Rodriguez-Rivera}, \citenamefont {Broholm},\ and\
  \citenamefont {Lee}}]{han_fractionalized_2012}%
  \BibitemOpen
  \bibfield  {author} {\bibinfo {author} {\bibfnamefont {T.~H.}\ \bibnamefont
  {Han}}, \bibinfo {author} {\bibfnamefont {J.~S.}\ \bibnamefont {Helton}},
  \bibinfo {author} {\bibfnamefont {S.}~\bibnamefont {Chu}}, \bibinfo {author}
  {\bibfnamefont {D.~G.}\ \bibnamefont {Nocera}}, \bibinfo {author}
  {\bibfnamefont {J.~A.}\ \bibnamefont {Rodriguez-Rivera}}, \bibinfo {author}
  {\bibfnamefont {C.}~\bibnamefont {Broholm}}, \ and\ \bibinfo {author}
  {\bibfnamefont {Y.~S.}\ \bibnamefont {Lee}},\ }\bibfield  {title}
  {{\selectlanguage {english}\enquote {\bibinfo {title} {{Fractionalized}
  excitations in the spin-liquid state of a kagome-lattice antiferromagnet},}\
  }}\href {\doibase 10.1038/nature11659} {\bibfield  {journal} {\bibinfo
  {journal} {Nature}\ }\textbf {\bibinfo {volume} {492}},\ \bibinfo {pages}
  {406--410} (\bibinfo {year} {2012})},\ \bibinfo {note} {number: 7429
  Publisher: Nature Publishing Group}\BibitemShut {NoStop}%
\bibitem [{\citenamefont {Li}\ \emph {et~al.}(2021{\natexlab{b}})\citenamefont
  {Li}, \citenamefont {Zhang}, \citenamefont {Yilmaz}, \citenamefont {Pai},
  \citenamefont {Marvinney}, \citenamefont {Said}, \citenamefont {Yin},
  \citenamefont {Gong}, \citenamefont {Tu}, \citenamefont {Vescovo},
  \citenamefont {Nelson}, \citenamefont {Moore}, \citenamefont {Murakami},
  \citenamefont {Lei}, \citenamefont {Lee}, \citenamefont {Lawrie},\ and\
  \citenamefont {Miao}}]{li_observation_2021}%
  \BibitemOpen
  \bibfield  {author} {\bibinfo {author} {\bibfnamefont {H.}~\bibnamefont
  {Li}}, \bibinfo {author} {\bibfnamefont {T.~T.}\ \bibnamefont {Zhang}},
  \bibinfo {author} {\bibfnamefont {T.}~\bibnamefont {Yilmaz}}, \bibinfo
  {author} {\bibfnamefont {Y.~Y.}\ \bibnamefont {Pai}}, \bibinfo {author}
  {\bibfnamefont {C.~E.}\ \bibnamefont {Marvinney}}, \bibinfo {author}
  {\bibfnamefont {A.}~\bibnamefont {Said}}, \bibinfo {author} {\bibfnamefont
  {Q.~W.}\ \bibnamefont {Yin}}, \bibinfo {author} {\bibfnamefont {C.~S.}\
  \bibnamefont {Gong}}, \bibinfo {author} {\bibfnamefont {Z.~J.}\ \bibnamefont
  {Tu}}, \bibinfo {author} {\bibfnamefont {E.}~\bibnamefont {Vescovo}},
  \bibinfo {author} {\bibfnamefont {C.~S.}\ \bibnamefont {Nelson}}, \bibinfo
  {author} {\bibfnamefont {R.~G.}\ \bibnamefont {Moore}}, \bibinfo {author}
  {\bibfnamefont {S.}~\bibnamefont {Murakami}}, \bibinfo {author}
  {\bibfnamefont {H.~C.}\ \bibnamefont {Lei}}, \bibinfo {author} {\bibfnamefont
  {H.~N.}\ \bibnamefont {Lee}}, \bibinfo {author} {\bibfnamefont {B.~J.}\
  \bibnamefont {Lawrie}}, \ and\ \bibinfo {author} {\bibfnamefont
  {H.}~\bibnamefont {Miao}},\ }\bibfield  {title} {\enquote {\bibinfo {title}
  {{Observation} of {Unconventional} {Charge} {Density} {Wave} without
  {Acoustic} {Phonon} {Anomaly} in {Kagome} {Superconductors}
  {AV$_{3}$Sb$_{5}$} {(A=Rb, Cs)}},}\ }\href {\doibase
  10.1103/PhysRevX.11.031050} {\bibfield  {journal} {\bibinfo  {journal} {Phys.
  Rev. X}\ }\textbf {\bibinfo {volume} {11}},\ \bibinfo {pages} {031050}
  (\bibinfo {year} {2021}{\natexlab{b}})},\ \bibinfo {note} {publisher:
  American Physical Society}\BibitemShut {NoStop}%
\bibitem [{\citenamefont {Tanaka}\ \emph {et~al.}(2020)\citenamefont {Tanaka},
  \citenamefont {Fujisawa}, \citenamefont {Kuroda}, \citenamefont {Noguchi},
  \citenamefont {Sakuragi}, \citenamefont {Bareille}, \citenamefont {Smith},
  \citenamefont {Cacho}, \citenamefont {Jung}, \citenamefont {Muro},
  \citenamefont {Okada},\ and\ \citenamefont
  {Kondo}}]{tanaka_three-dimensional_2020}%
  \BibitemOpen
  \bibfield  {author} {\bibinfo {author} {\bibfnamefont {H.}~\bibnamefont
  {Tanaka}}, \bibinfo {author} {\bibfnamefont {Y.}~\bibnamefont {Fujisawa}},
  \bibinfo {author} {\bibfnamefont {K.}~\bibnamefont {Kuroda}}, \bibinfo
  {author} {\bibfnamefont {R.}~\bibnamefont {Noguchi}}, \bibinfo {author}
  {\bibfnamefont {S.}~\bibnamefont {Sakuragi}}, \bibinfo {author}
  {\bibfnamefont {C.}~\bibnamefont {Bareille}}, \bibinfo {author}
  {\bibfnamefont {B.}~\bibnamefont {Smith}}, \bibinfo {author} {\bibfnamefont
  {C.}~\bibnamefont {Cacho}}, \bibinfo {author} {\bibfnamefont {S.~W.}\
  \bibnamefont {Jung}}, \bibinfo {author} {\bibfnamefont {T.}~\bibnamefont
  {Muro}}, \bibinfo {author} {\bibfnamefont {Y.}~\bibnamefont {Okada}}, \ and\
  \bibinfo {author} {\bibfnamefont {T.}~\bibnamefont {Kondo}},\ }\bibfield
  {title} {\enquote {\bibinfo {title} {{Three}-dimensional electronic structure
  in ferromagnetic {Fe$_{3}$Sn$_{2}$} with breathing kagome bilayers},}\ }\href
  {\doibase 10.1103/PhysRevB.101.161114} {\bibfield  {journal} {\bibinfo
  {journal} {Phys. Rev. B}\ }\textbf {\bibinfo {volume} {101}},\ \bibinfo
  {pages} {161114} (\bibinfo {year} {2020})},\ \bibinfo {note} {publisher:
  American Physical Society}\BibitemShut {NoStop}%
\bibitem [{\citenamefont {Yamaguchi}\ and\ \citenamefont
  {Watanabe}(1967)}]{yamaguchi_neutron_1967}%
  \BibitemOpen
  \bibfield  {author} {\bibinfo {author} {\bibfnamefont {K.}~\bibnamefont
  {Yamaguchi}}\ and\ \bibinfo {author} {\bibfnamefont {H.}~\bibnamefont
  {Watanabe}},\ }\bibfield  {title} {\enquote {\bibinfo {title} {{Neutron}
  {Diffraction} {Study} of {FeSn}},}\ }\href {\doibase 10.1143/JPSJ.22.1210}
  {\bibfield  {journal} {\bibinfo  {journal} {J. Phys. Soc. Jpn.}\ }\textbf
  {\bibinfo {volume} {22}},\ \bibinfo {pages} {1210--1213} (\bibinfo {year}
  {1967})},\ \bibinfo {note} {publisher: The Physical Society of
  Japan}\BibitemShut {NoStop}%
\bibitem [{\citenamefont {Häggström}\ \emph {et~al.}(1975)\citenamefont
  {Häggström}, \citenamefont {Ericsson}, \citenamefont {Wäppling},\ and\
  \citenamefont {Chandra}}]{haggstrom_studies_1975}%
  \BibitemOpen
  \bibfield  {author} {\bibinfo {author} {\bibfnamefont {L.}~\bibnamefont
  {Häggström}}, \bibinfo {author} {\bibfnamefont {T.}~\bibnamefont
  {Ericsson}}, \bibinfo {author} {\bibfnamefont {R.}~\bibnamefont {Wäppling}},
  \ and\ \bibinfo {author} {\bibfnamefont {K.}~\bibnamefont {Chandra}},\
  }\bibfield  {title} {{\selectlanguage {english}\enquote {\bibinfo {title}
  {Studies of the {Magnetic} {Structure} of {FeSn} {Using} the {Mössbauer}
  {Effect}},}\ }}\href {\doibase 10.1088/0031-8949/11/1/008} {\bibfield
  {journal} {\bibinfo  {journal} {Phys. Scr.}\ }\textbf {\bibinfo {volume}
  {11}},\ \bibinfo {pages} {47} (\bibinfo {year} {1975})}\BibitemShut {NoStop}%
\bibitem [{\citenamefont {Sales}\ \emph {et~al.}(2019)\citenamefont {Sales},
  \citenamefont {Yan}, \citenamefont {Meier}, \citenamefont {Christianson},
  \citenamefont {Okamoto},\ and\ \citenamefont
  {McGuire}}]{sales_electronic_2019}%
  \BibitemOpen
  \bibfield  {author} {\bibinfo {author} {\bibfnamefont {B.~C.}\ \bibnamefont
  {Sales}}, \bibinfo {author} {\bibfnamefont {J.}~\bibnamefont {Yan}}, \bibinfo
  {author} {\bibfnamefont {W.~R.}\ \bibnamefont {Meier}}, \bibinfo {author}
  {\bibfnamefont {A.~D.}\ \bibnamefont {Christianson}}, \bibinfo {author}
  {\bibfnamefont {S.}~\bibnamefont {Okamoto}}, \ and\ \bibinfo {author}
  {\bibfnamefont {M.~A.}\ \bibnamefont {McGuire}},\ }\bibfield  {title}
  {\enquote {\bibinfo {title} {{Electronic}, magnetic, and thermodynamic
  properties of the kagome layer compound {FeSn}},}\ }\href {\doibase
  10.1103/PhysRevMaterials.3.114203} {\bibfield  {journal} {\bibinfo  {journal}
  {Phys. Rev. Mater.}\ }\textbf {\bibinfo {volume} {3}},\ \bibinfo {pages}
  {114203} (\bibinfo {year} {2019})},\ \bibinfo {note} {publisher: American
  Physical Society}\BibitemShut {NoStop}%
\bibitem [{\citenamefont {Malaman}\ \emph {et~al.}(1978)\citenamefont
  {Malaman}, \citenamefont {Fruchart},\ and\ \citenamefont
  {Caer}}]{malaman_magnetic_1978}%
  \BibitemOpen
  \bibfield  {author} {\bibinfo {author} {\bibfnamefont {B.}~\bibnamefont
  {Malaman}}, \bibinfo {author} {\bibfnamefont {D.}~\bibnamefont {Fruchart}}, \
  and\ \bibinfo {author} {\bibfnamefont {G.~L.}\ \bibnamefont {Caer}},\
  }\bibfield  {title} {{\selectlanguage {english}\enquote {\bibinfo {title}
  {{Magnetic} properties of {Fe$_{3}$Sn$_{2}$}. {II}. {Neutron} diffraction
  study (and {Mossbauer} effect)},}\ }}\href {\doibase
  10.1088/0305-4608/8/11/022} {\bibfield  {journal} {\bibinfo  {journal} {J.
  Phys. F: Met. Phys.}\ }\textbf {\bibinfo {volume} {8}},\ \bibinfo {pages}
  {2389} (\bibinfo {year} {1978})}\BibitemShut {NoStop}%
\bibitem [{\citenamefont {Caer}\ \emph {et~al.}(1979)\citenamefont {Caer},
  \citenamefont {Malaman}, \citenamefont {Haggstrom},\ and\ \citenamefont
  {Ericsson}}]{caer_magnetic_1979}%
  \BibitemOpen
  \bibfield  {author} {\bibinfo {author} {\bibfnamefont {G.~L.}\ \bibnamefont
  {Caer}}, \bibinfo {author} {\bibfnamefont {B.}~\bibnamefont {Malaman}},
  \bibinfo {author} {\bibfnamefont {L.}~\bibnamefont {Haggstrom}}, \ and\
  \bibinfo {author} {\bibfnamefont {T.}~\bibnamefont {Ericsson}},\ }\bibfield
  {title} {{\selectlanguage {english}\enquote {\bibinfo {title} {Magnetic
  properties of {Fe$_{3}$Sn$_{2}$}. {III}. {A} {$^{119}$Sn} {Mossbauer}
  study},}\ }}\href {\doibase 10.1088/0305-4608/9/9/020} {\bibfield  {journal}
  {\bibinfo  {journal} {J. Phys. F: Met. Phys.}\ }\textbf {\bibinfo {volume}
  {9}},\ \bibinfo {pages} {1905} (\bibinfo {year} {1979})}\BibitemShut
  {NoStop}%
\bibitem [{\citenamefont {Fenner}\ \emph {et~al.}(2009)\citenamefont {Fenner},
  \citenamefont {Dee},\ and\ \citenamefont
  {Wills}}]{fenner_non-collinearity_2009}%
  \BibitemOpen
  \bibfield  {author} {\bibinfo {author} {\bibfnamefont {L.~A.}\ \bibnamefont
  {Fenner}}, \bibinfo {author} {\bibfnamefont {A.~A.}\ \bibnamefont {Dee}}, \
  and\ \bibinfo {author} {\bibfnamefont {A.~S.}\ \bibnamefont {Wills}},\
  }\bibfield  {title} {{\selectlanguage {english}\enquote {\bibinfo {title}
  {{Non}-collinearity and spin frustration in the itinerant kagome ferromagnet
  {Fe$_{3}$Sn$_{2}$}},}\ }}\href {\doibase 10.1088/0953-8984/21/45/452202}
  {\bibfield  {journal} {\bibinfo  {journal} {J. Phys.: Condens. Matter}\
  }\textbf {\bibinfo {volume} {21}},\ \bibinfo {pages} {452202} (\bibinfo
  {year} {2009})}\BibitemShut {NoStop}%
\bibitem [{\citenamefont {Dally}\ \emph {et~al.}(2021)\citenamefont {Dally},
  \citenamefont {Phelan}, \citenamefont {Bishop}, \citenamefont {Ghimire},\
  and\ \citenamefont {Lynn}}]{dally_isotropic_2021}%
  \BibitemOpen
  \bibfield  {author} {\bibinfo {author} {\bibfnamefont {R.~L.}\ \bibnamefont
  {Dally}}, \bibinfo {author} {\bibfnamefont {D.}~\bibnamefont {Phelan}},
  \bibinfo {author} {\bibfnamefont {N.}~\bibnamefont {Bishop}}, \bibinfo
  {author} {\bibfnamefont {N.~J.}\ \bibnamefont {Ghimire}}, \ and\ \bibinfo
  {author} {\bibfnamefont {J.~W.}\ \bibnamefont {Lynn}},\ }\bibfield  {title}
  {{\selectlanguage {english}\enquote {\bibinfo {title} {{Isotropic} {Nature}
  of the {Metallic} {Kagome} {Ferromagnet} {Fe$_{3}$Sn$_{2}$} at {High}
  {Temperatures}},}\ }}\href {\doibase 10.3390/cryst11030307} {\bibfield
  {journal} {\bibinfo  {journal} {Crystals}\ }\textbf {\bibinfo {volume}
  {11}},\ \bibinfo {pages} {307} (\bibinfo {year} {2021})},\ \bibinfo {note}
  {number: 3 Publisher: Multidisciplinary Digital Publishing
  Institute}\BibitemShut {NoStop}%
\bibitem [{\citenamefont {Lu}\ \emph {et~al.}(2013)\citenamefont {Lu},
  \citenamefont {Fu}, \citenamefont {Joannopoulos},\ and\ \citenamefont
  {Soljačić}}]{lu_weyl_2013}%
  \BibitemOpen
  \bibfield  {author} {\bibinfo {author} {\bibfnamefont {L.}~\bibnamefont
  {Lu}}, \bibinfo {author} {\bibfnamefont {L.}~\bibnamefont {Fu}}, \bibinfo
  {author} {\bibfnamefont {J.~D.}\ \bibnamefont {Joannopoulos}}, \ and\
  \bibinfo {author} {\bibfnamefont {M.}~\bibnamefont {Soljačić}},\ }\bibfield
   {title} {{\selectlanguage {english}\enquote {\bibinfo {title} {{Weyl} points
  and line nodes in gyroid photonic crystals},}\ }}\href {\doibase
  10.1038/nphoton.2013.42} {\bibfield  {journal} {\bibinfo  {journal} {Nat.
  Photon}\ }\textbf {\bibinfo {volume} {7}},\ \bibinfo {pages} {294--299}
  (\bibinfo {year} {2013})},\ \bibinfo {note} {number: 4 Publisher: Nature
  Publishing Group}\BibitemShut {NoStop}%
\bibitem [{\citenamefont {Yin}\ \emph {et~al.}(2019)\citenamefont {Yin},
  \citenamefont {Zhang}, \citenamefont {Chang}, \citenamefont {Wang},
  \citenamefont {Tsirkin}, \citenamefont {Guguchia}, \citenamefont {Lian},
  \citenamefont {Zhou}, \citenamefont {Jiang}, \citenamefont {Belopolski},
  \citenamefont {Shumiya}, \citenamefont {Multer}, \citenamefont {Litskevich},
  \citenamefont {Cochran}, \citenamefont {Lin}, \citenamefont {Wang},
  \citenamefont {Neupert}, \citenamefont {Jia}, \citenamefont {Lei},\ and\
  \citenamefont {Hasan}}]{yin_negative_2019}%
  \BibitemOpen
  \bibfield  {author} {\bibinfo {author} {\bibfnamefont {J.~X.}\ \bibnamefont
  {Yin}}, \bibinfo {author} {\bibfnamefont {S.~S.}\ \bibnamefont {Zhang}},
  \bibinfo {author} {\bibfnamefont {G.}~\bibnamefont {Chang}}, \bibinfo
  {author} {\bibfnamefont {Q.}~\bibnamefont {Wang}}, \bibinfo {author}
  {\bibfnamefont {S.~S.}\ \bibnamefont {Tsirkin}}, \bibinfo {author}
  {\bibfnamefont {Z.}~\bibnamefont {Guguchia}}, \bibinfo {author}
  {\bibfnamefont {B.}~\bibnamefont {Lian}}, \bibinfo {author} {\bibfnamefont
  {H.}~\bibnamefont {Zhou}}, \bibinfo {author} {\bibfnamefont {K.}~\bibnamefont
  {Jiang}}, \bibinfo {author} {\bibfnamefont {I.}~\bibnamefont {Belopolski}},
  \bibinfo {author} {\bibfnamefont {N.}~\bibnamefont {Shumiya}}, \bibinfo
  {author} {\bibfnamefont {D.}~\bibnamefont {Multer}}, \bibinfo {author}
  {\bibfnamefont {M.}~\bibnamefont {Litskevich}}, \bibinfo {author}
  {\bibfnamefont {T.~A.}\ \bibnamefont {Cochran}}, \bibinfo {author}
  {\bibfnamefont {H.}~\bibnamefont {Lin}}, \bibinfo {author} {\bibfnamefont
  {Z.}~\bibnamefont {Wang}}, \bibinfo {author} {\bibfnamefont {T.}~\bibnamefont
  {Neupert}}, \bibinfo {author} {\bibfnamefont {S.}~\bibnamefont {Jia}},
  \bibinfo {author} {\bibfnamefont {H.}~\bibnamefont {Lei}}, \ and\ \bibinfo
  {author} {\bibfnamefont {M.~Z.}\ \bibnamefont {Hasan}},\ }\bibfield  {title}
  {{\selectlanguage {english}\enquote {\bibinfo {title} {{Negative} flat band
  magnetism in a spin–orbit-coupled correlated kagome magnet},}\ }}\href
  {\doibase 10.1038/s41567-019-0426-7} {\bibfield  {journal} {\bibinfo
  {journal} {Nat. Phys.}\ }\textbf {\bibinfo {volume} {15}},\ \bibinfo {pages}
  {443--448} (\bibinfo {year} {2019})},\ \bibinfo {note} {number: 5 Publisher:
  Nature Publishing Group}\BibitemShut {NoStop}%
\bibitem [{\citenamefont {Xie}\ \emph {et~al.}(2022)\citenamefont {Xie},
  \citenamefont {Li}, \citenamefont {Bourges}, \citenamefont {Ivanov},
  \citenamefont {Ye}, \citenamefont {Yin}, \citenamefont {Hasan}, \citenamefont
  {Luo}, \citenamefont {Yao}, \citenamefont {Wang}, \citenamefont {Xu},\ and\
  \citenamefont {Dai}}]{xie_electron-phonon_2022}%
  \BibitemOpen
  \bibfield  {author} {\bibinfo {author} {\bibfnamefont {Y.}~\bibnamefont
  {Xie}}, \bibinfo {author} {\bibfnamefont {Y.}~\bibnamefont {Li}}, \bibinfo
  {author} {\bibfnamefont {P.}~\bibnamefont {Bourges}}, \bibinfo {author}
  {\bibfnamefont {A.}~\bibnamefont {Ivanov}}, \bibinfo {author} {\bibfnamefont
  {Z.}~\bibnamefont {Ye}}, \bibinfo {author} {\bibfnamefont {J.~X.}\
  \bibnamefont {Yin}}, \bibinfo {author} {\bibfnamefont {M.~Z.}\ \bibnamefont
  {Hasan}}, \bibinfo {author} {\bibfnamefont {A.}~\bibnamefont {Luo}}, \bibinfo
  {author} {\bibfnamefont {Y.}~\bibnamefont {Yao}}, \bibinfo {author}
  {\bibfnamefont {Z.}~\bibnamefont {Wang}}, \bibinfo {author} {\bibfnamefont
  {G.}~\bibnamefont {Xu}}, \ and\ \bibinfo {author} {\bibfnamefont
  {P.}~\bibnamefont {Dai}},\ }\bibfield  {title} {\enquote {\bibinfo {title}
  {{Electron}-phonon coupling in the charge density wave state of
  {CsV$_{3}$Sb$_{5}$}},}\ }\href {\doibase 10.1103/PhysRevB.105.L140501}
  {\bibfield  {journal} {\bibinfo  {journal} {Phys. Rev. B}\ }\textbf {\bibinfo
  {volume} {105}},\ \bibinfo {pages} {L140501} (\bibinfo {year} {2022})},\
  \bibinfo {note} {publisher: American Physical Society}\BibitemShut {NoStop}%
\bibitem [{\citenamefont {Do}\ \emph {et~al.}(2022)\citenamefont {Do},
  \citenamefont {Kaneko}, \citenamefont {Kajimoto}, \citenamefont {Kamazawa},
  \citenamefont {Stone}, \citenamefont {Lin}, \citenamefont {Itoh},
  \citenamefont {Masuda}, \citenamefont {Samolyuk}, \citenamefont {Dagotto},
  \citenamefont {Meier}, \citenamefont {Sales}, \citenamefont {Miao},\ and\
  \citenamefont {Christianson}}]{do_damped_2022}%
  \BibitemOpen
  \bibfield  {author} {\bibinfo {author} {\bibfnamefont {S.~H.}\ \bibnamefont
  {Do}}, \bibinfo {author} {\bibfnamefont {K.}~\bibnamefont {Kaneko}}, \bibinfo
  {author} {\bibfnamefont {R.}~\bibnamefont {Kajimoto}}, \bibinfo {author}
  {\bibfnamefont {K.}~\bibnamefont {Kamazawa}}, \bibinfo {author}
  {\bibfnamefont {M.~B.}\ \bibnamefont {Stone}}, \bibinfo {author}
  {\bibfnamefont {J.~Y.~Y.}\ \bibnamefont {Lin}}, \bibinfo {author}
  {\bibfnamefont {S.}~\bibnamefont {Itoh}}, \bibinfo {author} {\bibfnamefont
  {T.}~\bibnamefont {Masuda}}, \bibinfo {author} {\bibfnamefont {G.~D.}\
  \bibnamefont {Samolyuk}}, \bibinfo {author} {\bibfnamefont {E.}~\bibnamefont
  {Dagotto}}, \bibinfo {author} {\bibfnamefont {W.~R.}\ \bibnamefont {Meier}},
  \bibinfo {author} {\bibfnamefont {B.~C.}\ \bibnamefont {Sales}}, \bibinfo
  {author} {\bibfnamefont {H.}~\bibnamefont {Miao}}, \ and\ \bibinfo {author}
  {\bibfnamefont {A.~D.}\ \bibnamefont {Christianson}},\ }\bibfield  {title}
  {\enquote {\bibinfo {title} {Damped {Dirac} magnon in the metallic kagome
  antiferromagnet {FeSn}},}\ }\href {\doibase 10.1103/PhysRevB.105.L180403}
  {\bibfield  {journal} {\bibinfo  {journal} {Phys. Rev. B}\ }\textbf {\bibinfo
  {volume} {105}},\ \bibinfo {pages} {L180403} (\bibinfo {year} {2022})},\
  \bibinfo {note} {publisher: American Physical Society}\BibitemShut {NoStop}%
\bibitem [{\citenamefont {Xie}\ \emph {et~al.}(2021)\citenamefont {Xie},
  \citenamefont {Chen}, \citenamefont {Chen}, \citenamefont {Wang},
  \citenamefont {Yin}, \citenamefont {Stewart}, \citenamefont {Stone},
  \citenamefont {Daemen}, \citenamefont {Feng}, \citenamefont {Cao},
  \citenamefont {Lei}, \citenamefont {Yin}, \citenamefont {MacDonald},\ and\
  \citenamefont {Dai}}]{xie_spin_2021}%
  \BibitemOpen
  \bibfield  {author} {\bibinfo {author} {\bibfnamefont {Y.}~\bibnamefont
  {Xie}}, \bibinfo {author} {\bibfnamefont {L.}~\bibnamefont {Chen}}, \bibinfo
  {author} {\bibfnamefont {T.}~\bibnamefont {Chen}}, \bibinfo {author}
  {\bibfnamefont {Q.}~\bibnamefont {Wang}}, \bibinfo {author} {\bibfnamefont
  {Q.}~\bibnamefont {Yin}}, \bibinfo {author} {\bibfnamefont {J.~R.}\
  \bibnamefont {Stewart}}, \bibinfo {author} {\bibfnamefont {M.~B.}\
  \bibnamefont {Stone}}, \bibinfo {author} {\bibfnamefont {L.~L.}\ \bibnamefont
  {Daemen}}, \bibinfo {author} {\bibfnamefont {E.}~\bibnamefont {Feng}},
  \bibinfo {author} {\bibfnamefont {H.}~\bibnamefont {Cao}}, \bibinfo {author}
  {\bibfnamefont {H.}~\bibnamefont {Lei}}, \bibinfo {author} {\bibfnamefont
  {Z.}~\bibnamefont {Yin}}, \bibinfo {author} {\bibfnamefont {A.~H.}\
  \bibnamefont {MacDonald}}, \ and\ \bibinfo {author} {\bibfnamefont
  {P.}~\bibnamefont {Dai}},\ }\bibfield  {title} {{\selectlanguage
  {english}\enquote {\bibinfo {title} {Spin excitations in metallic kagome
  lattice {FeSn} and {CoSn}},}\ }}\href {\doibase 10.1038/s42005-021-00736-8}
  {\bibfield  {journal} {\bibinfo  {journal} {Commun. Phys.}\ }\textbf
  {\bibinfo {volume} {4}},\ \bibinfo {pages} {1--11} (\bibinfo {year}
  {2021})},\ \bibinfo {note} {number: 1 Publisher: Nature Publishing
  Group}\BibitemShut {NoStop}%
\bibitem [{\citenamefont {Wu}\ \emph {et~al.}(2021)\citenamefont {Wu},
  \citenamefont {Song}, \citenamefont {Yu}, \citenamefont {Wang}, \citenamefont
  {Xia}, \citenamefont {Hong}, \citenamefont {Zu}, \citenamefont {Du},
  \citenamefont {Vallobra}, \citenamefont {Liu}, \citenamefont {Torii},
  \citenamefont {Kamiyama}, \citenamefont {Xiong},\ and\ \citenamefont
  {Zhao}}]{wu_evidence_2021}%
  \BibitemOpen
  \bibfield  {author} {\bibinfo {author} {\bibfnamefont {P.}~\bibnamefont
  {Wu}}, \bibinfo {author} {\bibfnamefont {J.}~\bibnamefont {Song}}, \bibinfo
  {author} {\bibfnamefont {X.}~\bibnamefont {Yu}}, \bibinfo {author}
  {\bibfnamefont {Y.}~\bibnamefont {Wang}}, \bibinfo {author} {\bibfnamefont
  {K.}~\bibnamefont {Xia}}, \bibinfo {author} {\bibfnamefont {B.}~\bibnamefont
  {Hong}}, \bibinfo {author} {\bibfnamefont {L.}~\bibnamefont {Zu}}, \bibinfo
  {author} {\bibfnamefont {Y.}~\bibnamefont {Du}}, \bibinfo {author}
  {\bibfnamefont {P.}~\bibnamefont {Vallobra}}, \bibinfo {author}
  {\bibfnamefont {F.}~\bibnamefont {Liu}}, \bibinfo {author} {\bibfnamefont
  {S.}~\bibnamefont {Torii}}, \bibinfo {author} {\bibfnamefont
  {T.}~\bibnamefont {Kamiyama}}, \bibinfo {author} {\bibfnamefont
  {Y.}~\bibnamefont {Xiong}}, \ and\ \bibinfo {author} {\bibfnamefont
  {W.}~\bibnamefont {Zhao}},\ }\bibfield  {title} {\enquote {\bibinfo {title}
  {{Evidence} of spin reorientation and anharmonicity in kagome ferromagnet
  {Fe$_{3}$Sn$_{2}$}},}\ }\href {\doibase 10.1063/5.0063090} {\bibfield
  {journal} {\bibinfo  {journal} {Appl. Phys. Lett.}\ }\textbf {\bibinfo
  {volume} {119}},\ \bibinfo {pages} {082401} (\bibinfo {year} {2021})},\
  \bibinfo {note} {publisher: American Institute of Physics}\BibitemShut
  {NoStop}%
\bibitem [{\citenamefont {Paul}\ \emph {et~al.}(2020)\citenamefont {Paul},
  \citenamefont {Chung}, \citenamefont {Birol},\ and\ \citenamefont
  {Changlani}}]{paul_spin-lattice_2020}%
  \BibitemOpen
  \bibfield  {author} {\bibinfo {author} {\bibfnamefont {A.}~\bibnamefont
  {Paul}}, \bibinfo {author} {\bibfnamefont {C.~M.}\ \bibnamefont {Chung}},
  \bibinfo {author} {\bibfnamefont {T.}~\bibnamefont {Birol}}, \ and\ \bibinfo
  {author} {\bibfnamefont {H.~J.}\ \bibnamefont {Changlani}},\ }\bibfield
  {title} {\enquote {\bibinfo {title} {{Spin}-lattice {Coupling} and the
  {Emergence} of the {Trimerized} {Phase} in the {S}=1 {Kagome}
  {Antiferromagnet} {Na$_{2}$Ti$_{3}$Cl$_{8}$}},}\ }\href {\doibase
  10.1103/PhysRevLett.124.167203} {\bibfield  {journal} {\bibinfo  {journal}
  {Phys. Rev. Lett.}\ }\textbf {\bibinfo {volume} {124}},\ \bibinfo {pages}
  {167203} (\bibinfo {year} {2020})},\ \bibinfo {note} {publisher: American
  Physical Society}\BibitemShut {NoStop}%
\bibitem [{\citenamefont {Henriques}\ \emph {et~al.}(2016)\citenamefont
  {Henriques}, \citenamefont {Gorbunov}, \citenamefont {Kriegner},
  \citenamefont {Vališka}, \citenamefont {Andreev},\ and\ \citenamefont
  {Matěj}}]{henriques_magneto-elastic_2016}%
  \BibitemOpen
  \bibfield  {author} {\bibinfo {author} {\bibfnamefont {M.~S.}\ \bibnamefont
  {Henriques}}, \bibinfo {author} {\bibfnamefont {D.~I.}\ \bibnamefont
  {Gorbunov}}, \bibinfo {author} {\bibfnamefont {D.}~\bibnamefont {Kriegner}},
  \bibinfo {author} {\bibfnamefont {M.}~\bibnamefont {Vališka}}, \bibinfo
  {author} {\bibfnamefont {A.~V.}\ \bibnamefont {Andreev}}, \ and\ \bibinfo
  {author} {\bibfnamefont {Z.}~\bibnamefont {Matěj}},\ }\bibfield  {title}
  {{\selectlanguage {english}\enquote {\bibinfo {title} {{Magneto}-elastic
  coupling across the first-order transition in the distorted kagome lattice
  antiferromagnet {Dy$_{3}$Ru$_{4}$Al$_{12}$}},}\ }}\href {\doibase
  10.1016/j.jmmm.2015.07.066} {\bibfield  {journal} {\bibinfo  {journal} {J.
  Magn. Magn. Mater.}\ }\bibinfo {series} {Proceedings of the 20th
  {International} {Conference} on {Magnetism} ({Barcelona}) 5-10 {July} 2015},\
  \textbf {\bibinfo {volume} {400}},\ \bibinfo {pages} {125--129} (\bibinfo
  {year} {2016})}\BibitemShut {NoStop}%
\bibitem [{\citenamefont {Kresse}\ and\ \citenamefont
  {Furthmüller}(1996)}]{kresse_efficient_1996}%
  \BibitemOpen
  \bibfield  {author} {\bibinfo {author} {\bibfnamefont {G.}~\bibnamefont
  {Kresse}}\ and\ \bibinfo {author} {\bibfnamefont {J.}~\bibnamefont
  {Furthmüller}},\ }\bibfield  {title} {\enquote {\bibinfo {title}
  {{Efficient} iterative schemes for ab initio total-energy calculations using
  a plane-wave basis set},}\ }\href {\doibase 10.1103/PhysRevB.54.11169}
  {\bibfield  {journal} {\bibinfo  {journal} {Phys. Rev. B}\ }\textbf {\bibinfo
  {volume} {54}},\ \bibinfo {pages} {11169--11186} (\bibinfo {year} {1996})},\
  \bibinfo {note} {publisher: American Physical Society}\BibitemShut {NoStop}%
\bibitem [{\citenamefont {Blöchl}(1994)}]{blochl_projector_1994}%
  \BibitemOpen
  \bibfield  {author} {\bibinfo {author} {\bibfnamefont {P.~E.}\ \bibnamefont
  {Blöchl}},\ }\bibfield  {title} {\enquote {\bibinfo {title} {{Projector}
  augmented-wave method},}\ }\href {\doibase 10.1103/PhysRevB.50.17953}
  {\bibfield  {journal} {\bibinfo  {journal} {Phys. Rev. B}\ }\textbf {\bibinfo
  {volume} {50}},\ \bibinfo {pages} {17953--17979} (\bibinfo {year} {1994})},\
  \bibinfo {note} {publisher: American Physical Society}\BibitemShut {NoStop}%
\bibitem [{\citenamefont {Kresse}\ and\ \citenamefont
  {Joubert}(1999)}]{kresse_ultrasoft_1999}%
  \BibitemOpen
  \bibfield  {author} {\bibinfo {author} {\bibfnamefont {G.}~\bibnamefont
  {Kresse}}\ and\ \bibinfo {author} {\bibfnamefont {D.}~\bibnamefont
  {Joubert}},\ }\bibfield  {title} {\enquote {\bibinfo {title} {{From}
  ultrasoft pseudopotentials to the projector augmented-wave method},}\ }\href
  {\doibase 10.1103/PhysRevB.59.1758} {\bibfield  {journal} {\bibinfo
  {journal} {Phys. Rev. B}\ }\textbf {\bibinfo {volume} {59}},\ \bibinfo
  {pages} {1758--1775} (\bibinfo {year} {1999})},\ \bibinfo {note} {publisher:
  American Physical Society}\BibitemShut {NoStop}%
\bibitem [{\citenamefont {Perdew}\ \emph {et~al.}(1996)\citenamefont {Perdew},
  \citenamefont {Burke},\ and\ \citenamefont
  {Ernzerhof}}]{perdew_generalized_1996}%
  \BibitemOpen
  \bibfield  {author} {\bibinfo {author} {\bibfnamefont {J.~P.}\ \bibnamefont
  {Perdew}}, \bibinfo {author} {\bibfnamefont {K.}~\bibnamefont {Burke}}, \
  and\ \bibinfo {author} {\bibfnamefont {M.}~\bibnamefont {Ernzerhof}},\
  }\bibfield  {title} {\enquote {\bibinfo {title} {{Generalized} {Gradient}
  {Approximation} {Made} {Simple}},}\ }\href {\doibase
  10.1103/PhysRevLett.77.3865} {\bibfield  {journal} {\bibinfo  {journal}
  {Phys. Rev. Lett.}\ }\textbf {\bibinfo {volume} {77}},\ \bibinfo {pages}
  {3865--3868} (\bibinfo {year} {1996})},\ \bibinfo {note} {publisher: American
  Physical Society}\BibitemShut {NoStop}%
\bibitem [{\citenamefont {der Kraan}\ and\ \citenamefont
  {Buschow}(1986)}]{van_der_kraan_57fe_1986}%
  \BibitemOpen
  \bibfield  {author} {\bibinfo {author} {\bibfnamefont {A.~M.~Van}\
  \bibnamefont {der Kraan}}\ and\ \bibinfo {author} {\bibfnamefont {K.~H.~J.}\
  \bibnamefont {Buschow}},\ }\bibfield  {title} {{\selectlanguage
  {english}\enquote {\bibinfo {title} {{The} {$^{57}$Fe} {Mössbauer} isomer
  shift in intermetallic compounds of iron},}\ }}\href {\doibase
  10.1016/0378-4363(86)90492-4} {\bibfield  {journal} {\bibinfo  {journal}
  {Physica B+C}\ }\textbf {\bibinfo {volume} {138}},\ \bibinfo {pages} {55--62}
  (\bibinfo {year} {1986})}\BibitemShut {NoStop}%
\bibitem [{\citenamefont {Malaman}\ \emph {et~al.}(1976)\citenamefont
  {Malaman}, \citenamefont {Roques}, \citenamefont {Courtois},\ and\
  \citenamefont {Protas}}]{malaman_structure_1976}%
  \BibitemOpen
  \bibfield  {author} {\bibinfo {author} {\bibfnamefont {B.}~\bibnamefont
  {Malaman}}, \bibinfo {author} {\bibfnamefont {B.}~\bibnamefont {Roques}},
  \bibinfo {author} {\bibfnamefont {A.}~\bibnamefont {Courtois}}, \ and\
  \bibinfo {author} {\bibfnamefont {J.}~\bibnamefont {Protas}},\ }\bibfield
  {title} {\enquote {\bibinfo {title} {{Structure} cristalline du stannure de
  fer {Fe$_{3}$Sn$_{2}$}},}\ }\href {\doibase 10.1107/S0567740876005323}
  {\bibfield  {journal} {\bibinfo  {journal} {Acta. Crystallogr. B.}\ }\textbf
  {\bibinfo {volume} {32}},\ \bibinfo {pages} {1348--1351} (\bibinfo {year}
  {1976})},\ \bibinfo {note} {number: 5 Publisher: International Union of
  Crystallography}\BibitemShut {NoStop}%
\bibitem [{\citenamefont {Togo}\ and\ \citenamefont
  {Tanaka}(2015)}]{togo_first_2015}%
  \BibitemOpen
  \bibfield  {author} {\bibinfo {author} {\bibfnamefont {A.}~\bibnamefont
  {Togo}}\ and\ \bibinfo {author} {\bibfnamefont {I.}~\bibnamefont {Tanaka}},\
  }\bibfield  {title} {{\selectlanguage {english}\enquote {\bibinfo {title}
  {{First} principles phonon calculations in materials science},}\ }}\href
  {\doibase 10.1016/j.scriptamat.2015.07.021} {\bibfield  {journal} {\bibinfo
  {journal} {Scr. Mater.}\ }\textbf {\bibinfo {volume} {108}},\ \bibinfo
  {pages} {1--5} (\bibinfo {year} {2015})}\BibitemShut {NoStop}%
\bibitem [{\citenamefont {Cheng}\ \emph {et~al.}(2019)\citenamefont {Cheng},
  \citenamefont {Daemen}, \citenamefont {Kolesnikov},\ and\ \citenamefont
  {Ramirez-Cuesta}}]{cheng_simulation_2019}%
  \BibitemOpen
  \bibfield  {author} {\bibinfo {author} {\bibfnamefont {Y.~Q.}\ \bibnamefont
  {Cheng}}, \bibinfo {author} {\bibfnamefont {L.~L.}\ \bibnamefont {Daemen}},
  \bibinfo {author} {\bibfnamefont {A.~I.}\ \bibnamefont {Kolesnikov}}, \ and\
  \bibinfo {author} {\bibfnamefont {A.~J.}\ \bibnamefont {Ramirez-Cuesta}},\
  }\bibfield  {title} {\enquote {\bibinfo {title} {{Simulation} of {Inelastic}
  {Neutron} {Scattering} {Spectra} {Using} {OCLIMAX}},}\ }\href {\doibase
  10.1021/acs.jctc.8b01250} {\bibfield  {journal} {\bibinfo  {journal} {J.
  Chem. Theory Comput.}\ }\textbf {\bibinfo {volume} {15}},\ \bibinfo {pages}
  {1974--1982} (\bibinfo {year} {2019})},\ \bibinfo {note} {publisher: American
  Chemical Society}\BibitemShut {NoStop}%
\bibitem [{\citenamefont {Kanematsu}\ \emph {et~al.}(1960)\citenamefont
  {Kanematsu}, \citenamefont {Yasukōchi},\ and\ \citenamefont
  {Ohoyama}}]{kanematsu_antiferromagnetism_1960}%
  \BibitemOpen
  \bibfield  {author} {\bibinfo {author} {\bibfnamefont {K.}~\bibnamefont
  {Kanematsu}}, \bibinfo {author} {\bibfnamefont {K.}~\bibnamefont
  {Yasukōchi}}, \ and\ \bibinfo {author} {\bibfnamefont {T.}~\bibnamefont
  {Ohoyama}},\ }\bibfield  {title} {\enquote {\bibinfo {title}
  {{Antiferromagnetism} of {FeSn$_{2}$}},}\ }\href {\doibase
  10.1143/JPSJ.15.2358} {\bibfield  {journal} {\bibinfo  {journal} {J. Phys.
  Soc. Japan}\ }\textbf {\bibinfo {volume} {15}},\ \bibinfo {pages}
  {2358--2358} (\bibinfo {year} {1960})}\BibitemShut {NoStop}%
\bibitem [{\citenamefont {Li}\ \emph {et~al.}(2020)\citenamefont {Li},
  \citenamefont {Zhang}, \citenamefont {Liang}, \citenamefont {Ding},
  \citenamefont {Chen}, \citenamefont {Shen}, \citenamefont {Li}, \citenamefont
  {Liu}, \citenamefont {Xi}, \citenamefont {Wu}, \citenamefont {Yao},
  \citenamefont {Yang},\ and\ \citenamefont {Wang}}]{li_large_2020}%
  \BibitemOpen
  \bibfield  {author} {\bibinfo {author} {\bibfnamefont {H.}~\bibnamefont
  {Li}}, \bibinfo {author} {\bibfnamefont {B.}~\bibnamefont {Zhang}}, \bibinfo
  {author} {\bibfnamefont {J.}~\bibnamefont {Liang}}, \bibinfo {author}
  {\bibfnamefont {B.}~\bibnamefont {Ding}}, \bibinfo {author} {\bibfnamefont
  {J.}~\bibnamefont {Chen}}, \bibinfo {author} {\bibfnamefont {J.}~\bibnamefont
  {Shen}}, \bibinfo {author} {\bibfnamefont {Z.}~\bibnamefont {Li}}, \bibinfo
  {author} {\bibfnamefont {E.}~\bibnamefont {Liu}}, \bibinfo {author}
  {\bibfnamefont {X.}~\bibnamefont {Xi}}, \bibinfo {author} {\bibfnamefont
  {G.}~\bibnamefont {Wu}}, \bibinfo {author} {\bibfnamefont {Y.}~\bibnamefont
  {Yao}}, \bibinfo {author} {\bibfnamefont {H.}~\bibnamefont {Yang}}, \ and\
  \bibinfo {author} {\bibfnamefont {W.}~\bibnamefont {Wang}},\ }\bibfield
  {title} {\enquote {\bibinfo {title} {{Large} anomalous {Hall} effect in a
  hexagonal ferromagnetic {Fe$_{5}$Sn$_{3}$} single crystal},}\ }\href
  {\doibase 10.1103/PhysRevB.101.140409} {\bibfield  {journal} {\bibinfo
  {journal} {Phys. Rev. B}\ }\textbf {\bibinfo {volume} {101}},\ \bibinfo
  {pages} {140409} (\bibinfo {year} {2020})},\ \bibinfo {note} {publisher:
  American Physical Society}\BibitemShut {NoStop}%
\bibitem [{\citenamefont {Sales}\ \emph {et~al.}(2014)\citenamefont {Sales},
  \citenamefont {Saparov}, \citenamefont {McGuire}, \citenamefont {Singh},\
  and\ \citenamefont {Parker}}]{sales_ferromagnetism_2014}%
  \BibitemOpen
  \bibfield  {author} {\bibinfo {author} {\bibfnamefont {B.~C.}\ \bibnamefont
  {Sales}}, \bibinfo {author} {\bibfnamefont {B.}~\bibnamefont {Saparov}},
  \bibinfo {author} {\bibfnamefont {M.~A.}\ \bibnamefont {McGuire}}, \bibinfo
  {author} {\bibfnamefont {D.~J.}\ \bibnamefont {Singh}}, \ and\ \bibinfo
  {author} {\bibfnamefont {D.~S.}\ \bibnamefont {Parker}},\ }\bibfield  {title}
  {{\selectlanguage {english}\enquote {\bibinfo {title} {{Ferromagnetism} of
  {Fe$_{3}$Sn} and {Alloys}},}\ }}\href {\doibase 10.1038/srep07024} {\bibfield
   {journal} {\bibinfo  {journal} {Sci. Rep.}\ }\textbf {\bibinfo {volume}
  {4}},\ \bibinfo {pages} {7024} (\bibinfo {year} {2014})},\ \bibinfo {note}
  {number: 1 Publisher: Nature Publishing Group}\BibitemShut {NoStop}%
\bibitem [{\citenamefont {Giefers}\ and\ \citenamefont
  {Nicol}(2006)}]{giefers_high_2006}%
  \BibitemOpen
  \bibfield  {author} {\bibinfo {author} {\bibfnamefont {H.}~\bibnamefont
  {Giefers}}\ and\ \bibinfo {author} {\bibfnamefont {M.}~\bibnamefont
  {Nicol}},\ }\bibfield  {title} {{\selectlanguage {english}\enquote {\bibinfo
  {title} {{High} pressure {X}-ray diffraction study of all {Fe}–{Sn}
  intermetallic compounds and one {Fe}–{Sn} solid solution},}\ }}\href
  {\doibase 10.1016/j.jallcom.2005.11.061} {\bibfield  {journal} {\bibinfo
  {journal} {J. Alloys Compd.}\ }\textbf {\bibinfo {volume} {422}},\ \bibinfo
  {pages} {132--144} (\bibinfo {year} {2006})}\BibitemShut {NoStop}%
\bibitem [{\citenamefont {Toth}\ and\ \citenamefont
  {Lake}(2015)}]{toth_linear_2015}%
  \BibitemOpen
  \bibfield  {author} {\bibinfo {author} {\bibfnamefont {S.}~\bibnamefont
  {Toth}}\ and\ \bibinfo {author} {\bibfnamefont {B.}~\bibnamefont {Lake}},\
  }\bibfield  {title} {{\selectlanguage {english}\enquote {\bibinfo {title}
  {{Linear} spin wave theory for single-{Q} incommensurate magnetic
  structures},}\ }}\href {\doibase 10.1088/0953-8984/27/16/166002} {\bibfield
  {journal} {\bibinfo  {journal} {J. Phys.: Condens. Matter}\ }\textbf
  {\bibinfo {volume} {27}},\ \bibinfo {pages} {166002} (\bibinfo {year}
  {2015})},\ \bibinfo {note} {publisher: IOP Publishing}\BibitemShut {NoStop}%
\bibitem [{\citenamefont {Demmel}\ and\ \citenamefont
  {Chatterji}(2007)}]{demmel_persistent_2007}%
  \BibitemOpen
  \bibfield  {author} {\bibinfo {author} {\bibfnamefont {F.}~\bibnamefont
  {Demmel}}\ and\ \bibinfo {author} {\bibfnamefont {T.}~\bibnamefont
  {Chatterji}},\ }\bibfield  {title} {\enquote {\bibinfo {title} {{Persistent}
  spin waves above the {Néel} temperature in {YMnO$_{3}$}},}\ }\href {\doibase
  10.1103/PhysRevB.76.212402} {\bibfield  {journal} {\bibinfo  {journal} {Phys.
  Rev. B}\ }\textbf {\bibinfo {volume} {76}},\ \bibinfo {pages} {212402}
  (\bibinfo {year} {2007})},\ \bibinfo {note} {publisher: American Physical
  Society}\BibitemShut {NoStop}%
\bibitem [{\citenamefont {Ptok}\ \emph {et~al.}(2021)\citenamefont {Ptok},
  \citenamefont {Kobiałka}, \citenamefont {Sternik}, \citenamefont
  {Łażewski}, \citenamefont {Jochym}, \citenamefont {Oleś}, \citenamefont
  {Stankov},\ and\ \citenamefont {Piekarz}}]{ptok_chiral_2021}%
  \BibitemOpen
  \bibfield  {author} {\bibinfo {author} {\bibfnamefont {A.}~\bibnamefont
  {Ptok}}, \bibinfo {author} {\bibfnamefont {A.}~\bibnamefont {Kobiałka}},
  \bibinfo {author} {\bibfnamefont {M.}~\bibnamefont {Sternik}}, \bibinfo
  {author} {\bibfnamefont {J.}~\bibnamefont {Łażewski}}, \bibinfo {author}
  {\bibfnamefont {P.~T.}\ \bibnamefont {Jochym}}, \bibinfo {author}
  {\bibfnamefont {A.~M.}\ \bibnamefont {Oleś}}, \bibinfo {author}
  {\bibfnamefont {S.}~\bibnamefont {Stankov}}, \ and\ \bibinfo {author}
  {\bibfnamefont {P.}~\bibnamefont {Piekarz}},\ }\bibfield  {title} {\enquote
  {\bibinfo {title} {Chiral phonons in the honeycomb sublattice of layered
  {CoSn}-like compounds},}\ }\href {\doibase 10.1103/PhysRevB.104.054305}
  {\bibfield  {journal} {\bibinfo  {journal} {Phys. Rev. B}\ }\textbf {\bibinfo
  {volume} {104}},\ \bibinfo {pages} {054305} (\bibinfo {year} {2021})},\
  \bibinfo {note} {publisher: American Physical Society}\BibitemShut {NoStop}%
\bibitem [{\citenamefont {Schneeloch}\ \emph {et~al.}(2022)\citenamefont
  {Schneeloch}, \citenamefont {Tao}, \citenamefont {Cheng}, \citenamefont
  {Daemen}, \citenamefont {Xu}, \citenamefont {Zhang},\ and\ \citenamefont
  {Louca}}]{schneeloch_gapless_2022}%
  \BibitemOpen
  \bibfield  {author} {\bibinfo {author} {\bibfnamefont {J.~A.}\ \bibnamefont
  {Schneeloch}}, \bibinfo {author} {\bibfnamefont {Y.}~\bibnamefont {Tao}},
  \bibinfo {author} {\bibfnamefont {Y.}~\bibnamefont {Cheng}}, \bibinfo
  {author} {\bibfnamefont {L.}~\bibnamefont {Daemen}}, \bibinfo {author}
  {\bibfnamefont {G.}~\bibnamefont {Xu}}, \bibinfo {author} {\bibfnamefont
  {Q.}~\bibnamefont {Zhang}}, \ and\ \bibinfo {author} {\bibfnamefont
  {D.}~\bibnamefont {Louca}},\ }\bibfield  {title} {{\selectlanguage
  {english}\enquote {\bibinfo {title} {{Gapless} {Dirac} magnons in
  {CrCl$_{3}$}},}\ }}\href {\doibase 10.1038/s41535-022-00473-3} {\bibfield
  {journal} {\bibinfo  {journal} {npj Quantum Mater.}\ }\textbf {\bibinfo
  {volume} {7}},\ \bibinfo {pages} {1--7} (\bibinfo {year} {2022})},\ \bibinfo
  {note} {number: 1 Publisher: Nature Publishing Group}\BibitemShut {NoStop}%
\bibitem [{\citenamefont {He}\ \emph {et~al.}(2022)\citenamefont {He},
  \citenamefont {Peis}, \citenamefont {Stumberger}, \citenamefont {Prodan},
  \citenamefont {Tsurkan}, \citenamefont {Unglert}, \citenamefont {Chioncel},
  \citenamefont {Kézsmárki},\ and\ \citenamefont {Hackl}}]{he_phonon_2022}%
  \BibitemOpen
  \bibfield  {author} {\bibinfo {author} {\bibfnamefont {G.}~\bibnamefont
  {He}}, \bibinfo {author} {\bibfnamefont {L.}~\bibnamefont {Peis}}, \bibinfo
  {author} {\bibfnamefont {R.}~\bibnamefont {Stumberger}}, \bibinfo {author}
  {\bibfnamefont {L.}~\bibnamefont {Prodan}}, \bibinfo {author} {\bibfnamefont
  {V.}~\bibnamefont {Tsurkan}}, \bibinfo {author} {\bibfnamefont
  {N.}~\bibnamefont {Unglert}}, \bibinfo {author} {\bibfnamefont
  {L.}~\bibnamefont {Chioncel}}, \bibinfo {author} {\bibfnamefont
  {I.}~\bibnamefont {Kézsmárki}}, \ and\ \bibinfo {author} {\bibfnamefont
  {R.}~\bibnamefont {Hackl}},\ }\bibfield  {title} {{\selectlanguage
  {english}\enquote {\bibinfo {title} {{Phonon} {Anomalies} {Associated} with
  {Spin} {Reorientation} in the {Kagome} {Ferromagnet} {Fe$_{3}$Sn$_{2}$}},}\
  }}\href {\doibase 10.1002/pssb.202100169} {\bibfield  {journal} {\bibinfo
  {journal} {Phys. Status Solidi B}\ }\textbf {\bibinfo {volume} {259}},\
  \bibinfo {pages} {2100169} (\bibinfo {year} {2022})},\ \bibinfo {note}
  {\_eprint:
  https://onlinelibrary.wiley.com/doi/pdf/10.1002/pssb.202100169}\BibitemShut
  {NoStop}%
\end{thebibliography}%

\end{document}